\documentclass[12pt]{mn2e}
\usepackage{graphicx}
\usepackage{epsf}
\usepackage{lscape}
\usepackage{longtable}
\usepackage{rotating}
\usepackage{color}

\newcommand{\apj}{ApJ}
\newcommand{\apjl}{ApJ}
\newcommand{\apjs}{ApJS}
\newcommand{\mnras}{MNRAS}
\newcommand{\aj}{AJ}

\begin{document}
\topmargin -0.5in 

\title[New Star Forming Galaxies at $z\approx 7$]{New Star Forming Galaxies at $z\approx 7$ from WFC3 Imaging}

\author[Stephen M. Wilkins, et al.\ ]  
{
Stephen M. Wilkins$^{1}$\thanks{E-mail: stephen.wilkins@astro.ox.ac.uk}, Andrew J. Bunker$^{1}$, Silvio Lorenzoni$^{1}$, Joseph Caruana$^{1}$ \\
$^1$\,University of Oxford, Department of Physics, Denys Wilkinson Building, Keble Road, OX1 3RH, U.K. \\}
\maketitle

\begin{abstract}
The addition of Wide Field Camera 3 (WFC3) on the {\em Hubble Space Telescope} ({\em HST}) has led to a dramatic increase in our ability to study the $z>6$ Universe. The improvement in the near-infrared (NIR) sensitivity of WFC3 over previous instruments has enabled us to reach apparent magnitudes approaching $29$ (AB). This allows us to probe the rest-frame ultraviolet (UV) continuum, redshifted into the NIR at $z>6$. Taking advantage of the large optical depths of the intergalactic medium at this redshift, resulting in the Lyman-$\alpha$ break, we use a combination of WFC3 imaging and pre-existing Advanced Camera for Surveys (ACS) imaging to search for $z\approx 7$ galaxies over 4 fields in and around Great Observatories Origins Survey (GOODS) South. Our analysis reveals 29 new $z\approx 7$ star forming galaxy candidates in addition to 15 pre-existing candidates already discovered in these fields. The improved statistics from our doubling of the robust sample of $z$-drop candidates confirms the previously observed evolution of the bright end of the luminosity function. 
\end{abstract} 


\section{Introduction}

There has been great progress in recent years in discovering star-forming galaxies at
high redshift ($z>5$), either through the Lyman-$\alpha$ emission line or their
rest-frame UV continuum colours, which are strongly affected by absorption by
intervening hydrogen (the Gunn-Peterson effect at $z>6.3$ provides almost
complete absorption shortward of Lyman-$\alpha$, e.g. Gunn \& Peterson 1965, Fan et al. 2001). This Lyman-break technique
was initially used to identify $z\approx 3$ galaxies from broad-band images
(Steidel et al.\ 1996), and improvements in sensitivity of near-infrared imaging have
enabled this to be pushed to higher redshifts (e.g. at $z\approx 6$ using the Advanced Camera for Surveys on {\em HST},
Stanway, Bunker \& McMahon 2003). 

Recently, the new {\em Wide Field Camera 3} (WFC3) camera has been installed on {\em HST}, which
has a near-infrared channel with significantly larger field and better sensitivity than the previous-generation {\em Near Infrared Camera and Multi-Object Spectrometer} (NICMOS) instrument. A number of multi-waveband imaging campaigns are underway, typically using broad-band filters at 1.0, 1.25 and 1.6$\mu$m and targetting fields with existing deep optical data. This is an ideal strategy to look for optical ``drop-outs'', seen only at the longer WFC3 wavelengths, which are candidate $z>7$ galaxies. A number of papers (e.g. Oesch et al. 2010, Bouwens et al. 2010, Bunker et al. 2010, McLure et al. 2010, Yan et al. 2010 and Finkelstein et al. 2010, Wilkins et al. 2010) have appeared in the past year presenting first results from such imaging. 

In this paper we analyse all the data taken so far in two large programs targetting fields in the vicinity of the Great Observatories Orgins Deep Survey-South (GOODS-South), which in one case includes the {\em Hubble} Ultra Deep Field, and present a new sample of $z$-band drop-outs which more than doubles the existing number of robust $z>7$ candidates. By building such large samples ($>40$ objects spanning 3 magnitudes) we can ultimately study the rest-UV luminosity function, and hence estimate the integrated star formation rate within 800\,Myr of the Big Bang. Understanding the global star formation history at these redshifts is crucial in answering the question whether the UV photons produced by the short-lived massive OB stars were sufficient to reionize the Universe.

This paper is organised as follows: in Section 2 we discuss the observations, our data reduction process and our photometry measurement and catalogue production technique. In Section 3 our candidate selection procedure is described and we present a list of $z$-drop candidates together with a comparison with other studies. In Section 4 we investigate the properties of these objects, including their redshift distribution, surface density, and luminosity function. Further, in this section we investigate the implications of the luminosity function for the star formation rate density and production of ionising photons. In Section 5 we present our conclusions. Throughout, we adopt the standard ‘concordance’ cosmology of $\Omega_{M}= 0.3$, $\Omega_{\Lambda}= 0.7$ and use $H_{0}=70$ kms$^{-1}$ Mpc$^{-1}$. All magnitudes are on the AB system (Oke \& Gunn 1983).

\section{Observations and Data Reduction}

\subsection{Observations}
In this paper we analyse ACS $b_{435w}$, $v_{606w}$, $i_{775w}$ and $z_{850lp}$-band imaging together with recently obtained near-infrared images from WFC3 on {\em HST} in $Y_{098m}/Y_{105w}$, $J_{125w}$ and $H_{160w}$-bands (see Figure 1, 1--1.6$\mu$m) taken in the vicinity of CDFS/GOODS-South.

The WFC3 data come from two different {\em HST} programs. The Early Release Science (ERS) program GO/DD\#11359
(P.I.\ R.~O'Connell) covers ten overlapping pointings with two orbits in each filter; analysis of the first 6 pointings
was presented in Wilkins et al.\ (2010), and we here we analyse the complete dataset (referred to as the ERS field). Secondly, deep WFC3 imaging of the {\em Hubble} Ultra Deep Field (HUDF, Beckwith et al. 2006) and two nearby deep flanking fields come from the {\em HST} Treasury programme GO-11563 (P.I.\ G.~Illingworth). These flanking fields were imaged by ACS in $v_{606w}$-, $i_{775w}$- and $z_{850lp}$-bands during 2005--6 in parallel with deep {\em HST} NICMOS NIC3 observations of the original UDF as part of program GO-10632 (P.I.\ M.~Stiavelli), and we adopt the original naming convention from the UDF parallels of HUDF (NIC)P12 and HUDF (NIC)P34 (as in Oesch et al.\ 2007, 2009), which we abbreviate to P12 and P34. We note that P12 is called HUDF05-01 in the WFC3 {\em HST} programme GO-11563, and P34 is HUDF05-02. The three WFC3 fields observed fall entirely within the $11\,{\mathrm arcmin}^{2}$ footprint of the ACS HUDF and flanking field images. This 192-orbit WFC3 program is ongoing, with the HUDF observations comprising  half of the total orbits. In Bunker et al.\ (2010) we analysed the HUDF data obtained soon-after the commissioning of WFC3, and in this paper we study the new images of the two deep flanking fields, and reanalyse the HUDF data using more recent on-orbit calibration of the WFC3 IR detector. 

The infrared channel of WFC3 was used, which is a Teledyne $1014\times 1014$ pixel HgCdTe detector,
with a field of view of $123"\times 136"$ (a 10-pixel strip on the edge is not illuminated by sky and used for pedestal estimation).
Both programs use the same $J$ and $H$-band filters (F125W and F160W), although the ERS images use a $Y$-band filter
(F098M) which covers only the blue side of the wider F105W filter used in the HUDF and flanking field images (see Figure 1).
The data were taken in ``MULTIACCUM" mode using SPARSAMPLE100, which non-destructively reads the array every 100\,seconds. These repeated non-destructive reads of the infrared array allow gradient-fitting to obtain the count rate (``sampling up the ramp'') and the flagging and rejection of cosmic ray strikes. 

For the HUDF and parallel fields, each MULTIACCUM exposure comprised 16 reads for a total duration of 1403\,sec per exposure, with two exposures per orbit. The WFC3 HUDF data, taken over 26 August -- 05 September 2009 U.T.,  comprises 39.2\,ksec (20 exposures) in $Y_{105w}$ (not including 8 exposures affected by severe persistence, Bunker et al.\ 2010), 44.8\,ksec (32 exposures) in $J_{125w}$ and 78.5\,ksec (54 exposures) in $H_{160w}$. The new P34 data (spanning 23 January -- 17 February 2010 U.T.) goes almost as deep, with 20, 28 \& 34 exposures in $Y_{105w}$, $J_{125w}$ \& $H_{160w}$ respectively (28.0\,ksec, 39.2\,ksec \& 47.7\,ksec); observations of this fields are nearly complete, unlike P12 where the bulk of the data will come next year. In this paper we analyse the existing 12 exposures (16.8\,ksec) in $Y_{105w}$-band and 24 exposures (33.6\,ksec) in $J_{125w}$-band for P12 (taken over 02 -- 15 November 2009 U.T.), although only 4 exposures (5.6\,ksec) have been taken so far in the $H_{160w}$-band. For the ERS data, each MULTIACCUM exposure comprised 9 or 10 non-destructive
read-outs, totalling 803--903\,s, and there were three such exposures per orbit (and each filter was imaged for 2 orbits per pointing,
with the 6 exposures totalling 2.6\,ksec). 

Advanced Camera for Surveys (ACS) optical/NIR ($0.4-1.0\mu m$) imaging for the 4 fields comes from a variety of sources. For the ERS field we utilise $b_{435w}$, $v_{606w}$, $i_{775w}$, and $z_{850lp}$-band imaging from GOODSv2.0 (Giavalisco et al. 2004). ACS $b_{435w}$, $v_{606w}$, $i_{775w}$, and $z_{850lp}$-band Imaging of the HUDF is taken from Beckwith et al.\ 2006. For the P12 field we use the publicly-available $v_{606w}, i_{775w}, z_{850lp}$ reductions provided by the UDF05 team (Oesch et al.\ 2007) and for the P34 field we obtain $v_{606w}, i_{775w}, z_{850lp}$-band imaging from the {\em HST} archive and reduce it as described below.

\subsection{Data Reduction}

\begin{table*}
\begin{tabular}{lccccccccc}
& & \multicolumn{7}{|c|}{ACS/WFC3 $2\sigma$ Detection Limit} \\
Field ID & Centre (J2000) & Area & $b_{435w}$&$v_{606w}$&$i_{775w}$ & $z_{850lp}$ & $Y_{098m/105w}$$^{a}$ & $J_{125w}$ ($7\sigma$) & $H_{160w}$ \\
\hline\hline
HUDF & 03:32:38.4 -27:47:00 & 4.2 &30.3 &30.7 &30.6 &30.0 & 29.65 &29.72 (28.36) & 29.67 \\
P34  & 3:33:05.3 -27:51:23 & 4.2 & - & 29.9 & 29.5 &29.7 &29.45 &29.58 (28.22) & 29.41\\
P12  & 3:33:01.9 -27:41:10 & 4.2 & - & 29.9 & 29.6 & 29.6 & 29.16 & 29.50 (28.14) & 28.23 \\
ERS & 3:32:23.6 -27:42:50 & 37.0  & 29.1 & 29.1 & 28.5 & 28.4 & 28.02 & 28.39 (27.03) & 28.10 \\

\end{tabular}
\caption{Summary of observations. $^{a}$$Y_{098m}$ for the ERS field and $Y_{105w}$ for the HUDF/P12/P34 fields measured in 0\farcs6 diameter apertures.}
\label{tab:obssum}
\end{table*}

The WFC3 imaging was reduced by first utilising the IRAF.STSDAS pipeline {\tt calwfc3} to calculate the count rate and reject cosmic rays through gradient fitting, as well as subtracting the zeroth read and flat-fielding. We used {\sc MULTIDRIZZLE} (Koekemoer et al.\ 2002) to combine exposures taken through the same filter in each pointing, taking account of the geometric distortions and mapping on to an output pixel size of $0\farcs06$ from an original $0\farcs13\,{\rm pix}^{-1}$. This was the same scale as we used in our analysis of the {\em Hubble} Ultra Deep Field WFC3 images (Bunker et al.\ 2010) and corresponds to a $2\times 2$ block-averaging of the GOODSv2.0 ACS drizzled images and HUDF ACS reduced images. 

The new data were reduced using the latest version of {\tt calwfc3} (29 October 2009
release), and we also re-reduced the HUDF images originally presented in Bunker et al.\ (2010) -- the
earlier paper had used a reduction using {\tt XDIMSUM} and fits to the geometric distortion due
to {\sc MULTIDRIZZLE} not then being available for the newly-commissioned WFC3. For the HUDF and flanking
fields, we used a {\sc DRIZZLE} pixel fraction of 0.6 to recover some of the under-sampling. In each of these three fields
we survey 4.18\,arcmin$^2$ in all exposures, with another 0.67\,arcmin$^2$ surveyed at half the maximum depth.
For the ERS field, we reduced the data for all 10 ERS pointings using the same technique. However, we used a pixel fraction of 1.0 in multidrizzle,
as we only had 6 exposures. We then mosaicked together the 10 ERS pointings in each filter, using inverse-variance weighting
for the overlap regions, producing a field of fairly uniform depth covering 37\,arcmin$^{2}$, with a further 8\,arcmin$^2$ going
less deep.

ACS $v_{606w}, i_{775w}, z_{850lp}$-band imaging of the P34 field taken from the {\em HST} archive was reduced using MULTIDRIZZLE to combine blocks of data taken close in time with similar telescope roll angles, using an output $0\farcs03$ pixel scale. These subsets of drizzled images were then registered and combined with {\tt IRAF.imcombine}. Our combined images were 4.8\,ksec in $v$-band, 10.6\,ksec in $i$-band and 26.8\,ksec in $z$-band. All the {\em ACS} images were then block-averaged $2\times 2$ and registered with our drizzled {\em WFC3} frames.

\subsubsection{Photometry and Zeropoints}

In our final combined $Y_{098m/105w}$-band images, we measure a FWHM of $\sim 0\farcs15$ for point sources in the field. As most high-redshift galaxies are likely to be barely resolved (e.g., Bunker et al.\ 2004, Fergusson et al.\ 2004) we perform photometry using fixed apertures of $0\farcs6$ diameter, and introduce an aperture correction to account for the flux falling outside of the aperture. This correction was determined to be 0.2\,mag in the $Y_{098m/125w}$-band  and 0.25\,mag in $J_{125w}$-band and $H_{160w}$-bands from photometry with larger apertures on bright but unsaturated point sources. We note that the $H_{160w}$-band images display significant Airy diffraction rings around point sources. For the ACS images, the better resolution and finer pixel sampling require a smaller aperture correction of $\approx 0.1$\,mag. 

For the ACS images we take the standard ACS zeropoints. For WFC3, we use the recent zeropoints reported on {\tt http://www.stsci.edu/hst/wfc3/phot\_zp\_lbn} during February 2010, where the $Y_{098m}$-band has an AB magnitude zeropoint of 25.68 (such that a source of this brightness would have a count rate of 1 electron per second), and zeropoints of 26.27, 26.25 \& 25.96 for $Y_{105w}$, $J_{125w}$ \& $H_{160w}$. We note that the information in the image headers of the earlier images (released in September 2009) is slightly different by 0.1-0.15\,mag, with zeropoints of $Y_{105w}=26.16$, $J_{125w}=26.10$ \& $H_{160w}=25.81$ (as used in Bunker et al.\ 2010).

We also apply a correction for the small amount of foreground Galactic extinction toward the GOODS-South using the {\it COBE}/DIRBE \& {\it IRAS}/ISSA dust maps of Schlegel, Finkbeiner \& Davis (1998). The optical reddening is $E(B-V)=0.009$, equivalent to extinctions of $A_{850lp}=0.012$,  $A_{105w}=0.010$, $A_{125w}=0.008$ \& $A_{160w}=0.005$.

Information, including the position, area, and average $2\sigma$ depths reached in each filter, for each field are given in Table 1.

\subsection{Construction of Catalogs}
\label{sec:catalos}

Candidate selection for all objects in the field was performed using version 2.5.0 of the SExtractor photometry package (Bertin \& Arnouts 1996). For the $z_{850lp}$-drops, as we are searching specifically for objects which are securely detected in the WFC3 $Y_{098m/105w}$-band, with minimal flux in the ACS images (see following section), fixed circular apertures $0\farcs6$ in diameter were `trained' in the $Y_{098m/105w}$-image and the identified apertures used to measure the flux at the same spatial location in the remaining ACS and WFC3 bands by running SExtractor in dual-image mode. For the initial object identification, we adopted a limit of at least 5 contiguous pixels above a threshold of $2\sigma$ per pixel (on the data drizzled to a scale of 0\farcs06~pixel$^{-1}$).

\section{Candidate Selection}

\begin{figure}
\centering
\includegraphics[width=18pc]{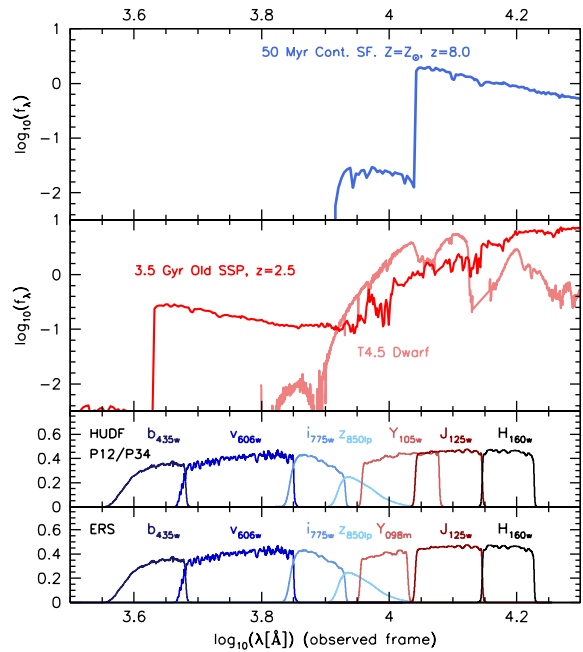}
\caption{Top panel - Model (from the Starburst99, Leitherer et al. 1999) spectral energy distribution (SED) of a redshifted $z=7$ star forming galaxy. Middle panel - Potential contaminants: Observed SED of a low-mass dwarf star (class: $T4.5$, Knapp et al. 2004) together with the model (Starburst99) SED of a $3.5$Gyr Single-aged Stellar Population at $z=1.8$. The bottom two panels show the transmission functions of the combination of filters available to each field.}
\label{fig:lbg_spectra}
\end{figure}

At very high redshifts the integrated optical depth of neutral hydrogen in the intergalactic medium (IGM) becomes large, resulting in a Gunn-Peterson trough (Gunn \& Peterson 1965), a large flux decrement shortward of the rest-frame wavelength of the Lyman-$\alpha$ transition ($1216{\rm \AA}$). At $z\approx 7$ the location of the Lyman-$\alpha$ break is redshifted to $\sim 1\mu m$ -- the ACS $z_{850lp}$ and WFC3 $Y_{105w/098m}$ are suitably located such that a $z=6.5-8.0$ star forming galaxy will experience a significant flux decrement between the two filters (see Figure \ref{fig:lbg_spectra}). Specifically a $z_{850lp}-Y_{105w}>1.0$ colour is indicative of such a spectral break.

In order to select only robust $z\approx 7$ star forming galaxy candidates we first introduce a $J_{125w}$ magnitude cut. The choice of the $J_{125w}$ filter for this cut is motivated by the fact that for $z\approx 7$ galaxies this filter will be uncontaminated by either the break or Lyman-$\alpha$ emission and will thus be an indication of the rest-frame ultraviolet luminosity of the candidates. For each field we define the $J_{125w}$ limit as the magnitude where the mean signal-to-noise ratio of an object at the limit is $7\sigma$.

\subsection{Contamination}

When searching for distant galaxies using only broadband photometry, contamination becomes a serious issue. Sources of contamination can be placed into three broad categories: objects whose intrinsic colours are similar to those of the target population; objects whose observed colours are similar because of photometric noise; and the effect of transient phenomena. We discuss each of these below.

\subsubsection{Intrinsically Red Objects}

Two distinct types of contaminating objects can have apparent $z_{850lp}-Y_{105w/098m}$ colours consistent with those of $z\approx 7$ galaxies, making them indistinguishable without the aid of additional constraints. These include lower-redshift galaxies where the $z_{850lp}$ and $Y_{105w/098m}$ filters straddle the Balmer/$4000{\rm \AA}$ break feature (at approximately $z\approx 1.5-1.9$) and low-mass dwarf stars, particularly those of the L and T classification whose low temperature and broad absorption features can mimic a spectral break.

Examples of the spectral energy distributions (SEDs) of each of these types of object (a model $3.5$ Gyr old single-aged stellar population at $z=1.8$ and a T4.5 dwarf star) are shown in Figure \ref{fig:lbg_spectra}. In both these cases the slope of the SED longward of the spectral break (i.e. longward of $z_{850lp}$) is somewhat redder than that predicted for a high-$z$ star forming galaxy. The addition of a further filter at wavelengths redder than the $Y_{105w/098m}$ filter can then be used to discriminate between high-$z$ galaxies and these lower-redshift interlopers.

\begin{figure}
\centering
\includegraphics[width=16pc]{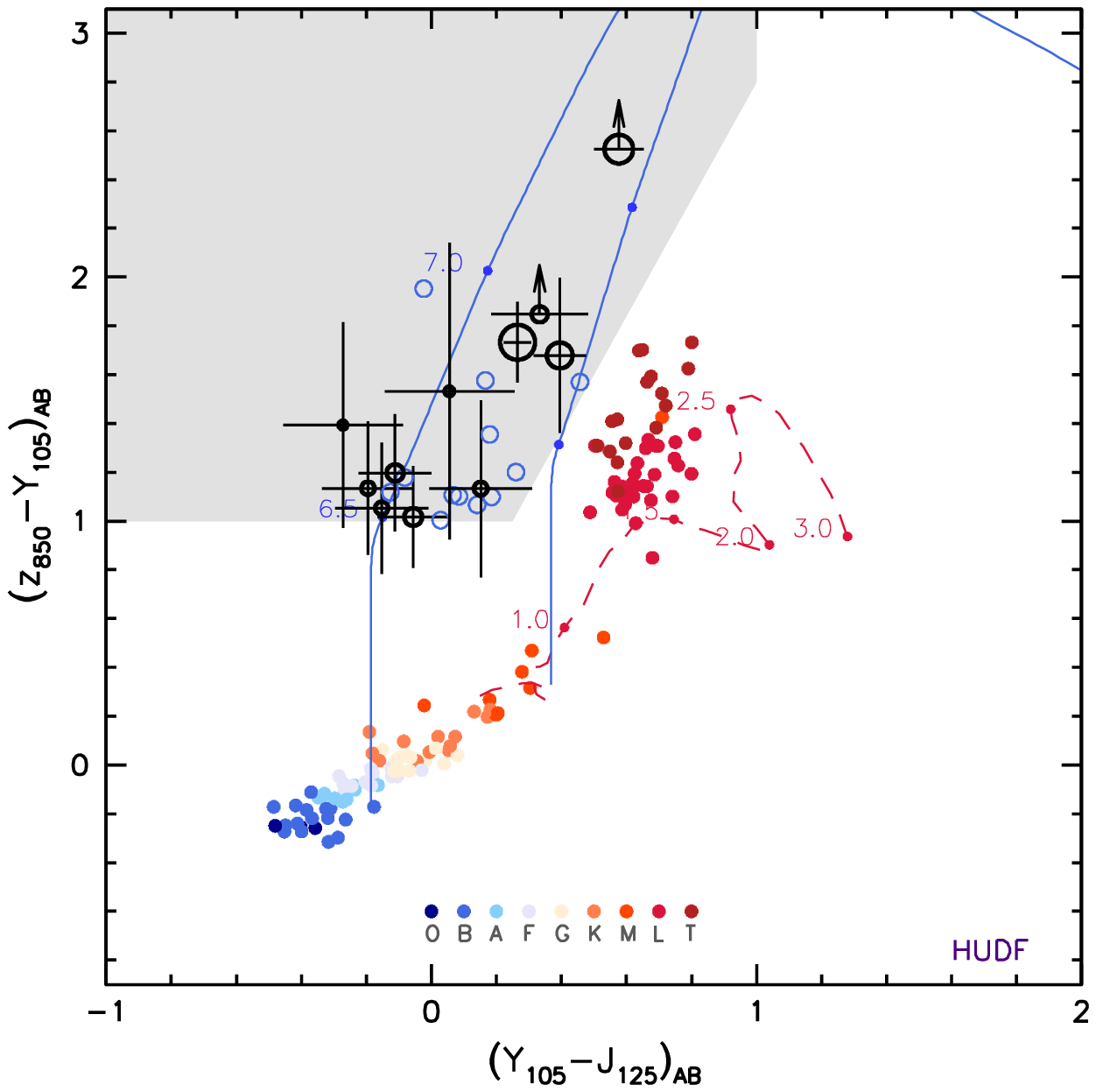}
\includegraphics[width=16pc]{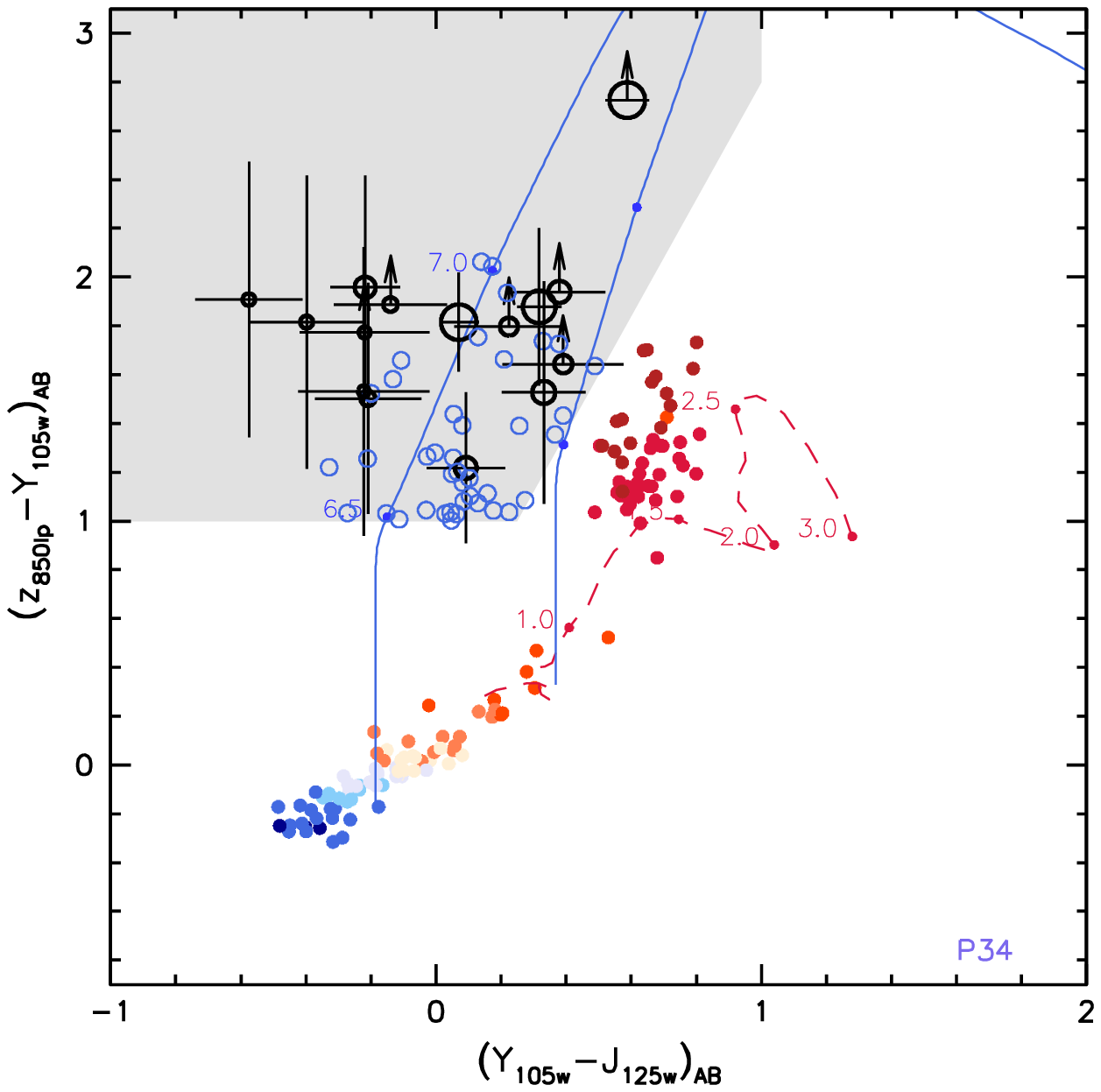}
\includegraphics[width=16pc]{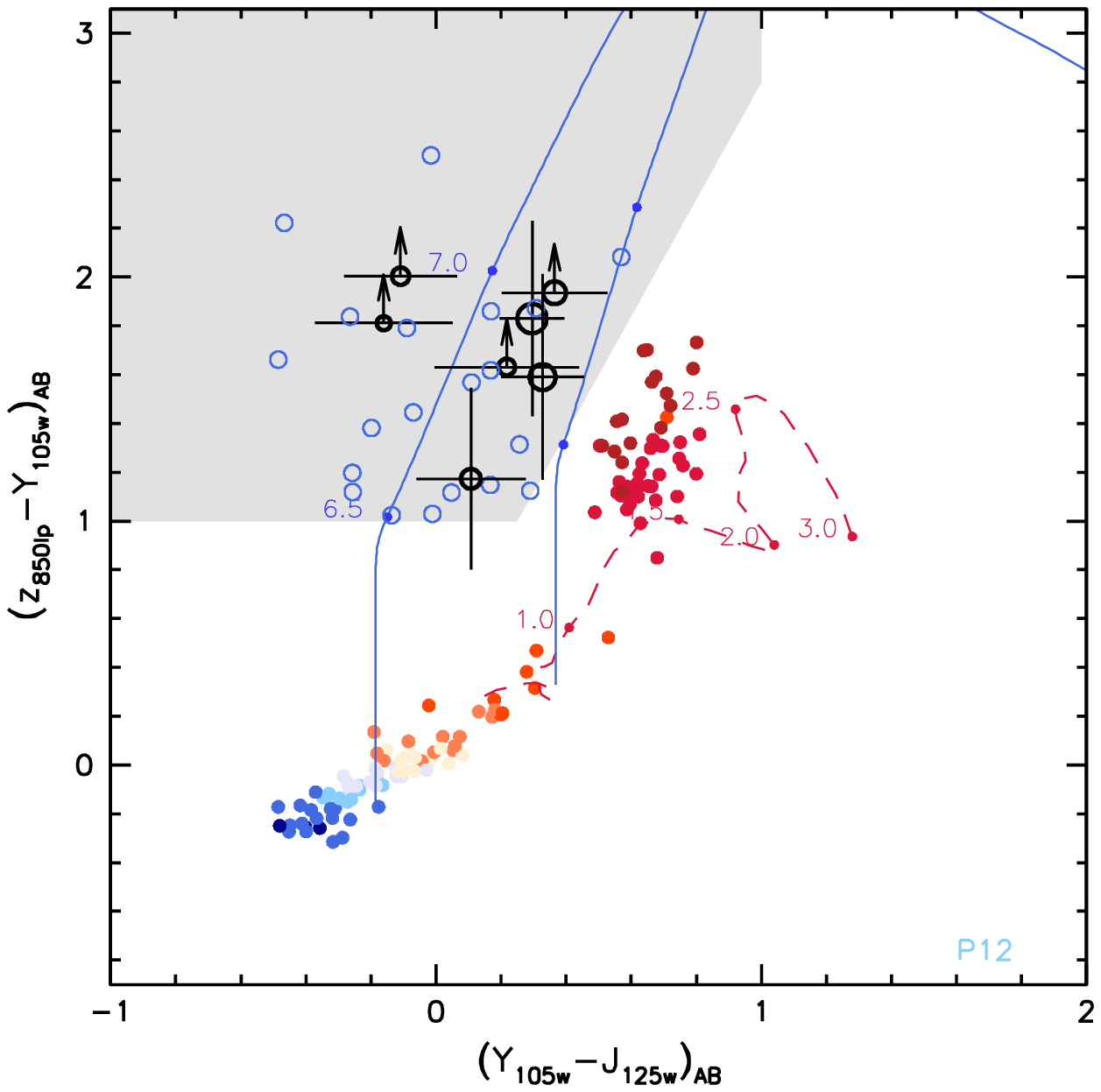}
\caption{$Y_{105w}-J_{125w}$ and $z_{850lp}-Y_{105w}$ colour - colour figures for the HUDF/P34/P12 fields showing our $zYJ$ colour selection window (grey shaded area), the location of our candidates, the predicted paths taken by high-$z$ galaxies (blue lines, $\beta=-3.0$ (left) and $\beta=0.0$ (right)) and the location of possible contaminating sources. Contaminating sources include galactic stars (denoted by coloured filled circles) and a passively evolving instantaneous burst of star formation that occured at $z=10$ (red dashed line). High-$z$ candidates are denoted by black circles (where the size of the circle is an indication of the apparent $J_{125w}$ magnitude). Small blue circles denote objects that meet our $zYJ$ colour criteria but which are detected in the bluer $(b)vi$ ACS bands.}
\label{fig:cc_1}
\end{figure}

\begin{figure}
\centering
\includegraphics[width=16pc]{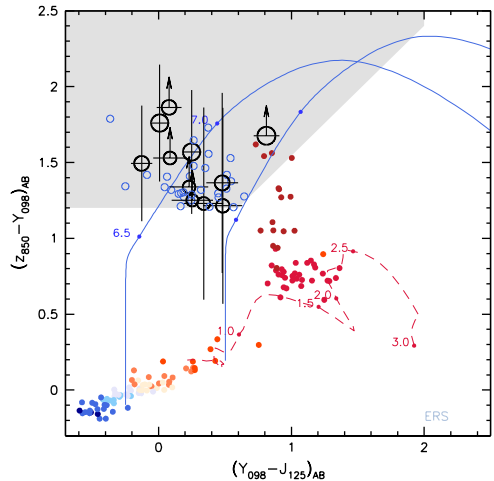}
\caption{$Y_{098m}-J_{125w}$ and $z_{850lp}-Y_{098m}$ colour - colour figures for the ERS field showing our $zYJ$ colour selection window (grey shaded area), the location of our candidates, the predicted paths taken by high-$z$ galaxies (blue lines, $\beta=-3.0$ (left) and $\beta=0.0$ (right)) and the location of possible contaminating sources. Contaminating sources include galactic stars (denoted by coloured filled circles) and a passively evolving instantaneous burst of star formation that occured at $z=10$ (red dashed line). High-$z$ candidates are denoted by black circles (where the size of the circle is an indication of the apparent $J_{125w}$ magnitude). Small blue circles denote objects that meet our $zYJ$ colour criteria but which are detected in the bluer $bvi$ ACS bands.}
\label{fig:cc_2}
\end{figure}

The ability for an additional redward filter to discriminate against these interlopers can be seen more clearly in Figures \ref{fig:cc_1} and \ref{fig:cc_2}. In these figures the positions of both the interlopers and the tracks expected for high-$z$ star forming galaxies are shown in the $(z_{850lp} - Y_{105w/098m})$ -- $(Y_{105w/098m} - J_{125w})$ colour plane. With the exception of the lowest temperature T dwarfs where the $Y_{098m}$ filter is employed (the ERS field) these interlopers form a distinct locus separate from $z\approx 7$ star forming galaxies with UV spectral slope index $\beta<0.0$ (where $f_{\lambda}=\lambda^{\beta}$ is used as a model of the UV properties of star forming galaxies). 

Using this analysis it is possible to design a window in $(z_{850lp} - Y_{105w/098m})$ -  $(Y_{105w/098m} - J_{125w})$ colour - colour space that selects only high-$z$ star forming galaxies. For the HUDF/P12/P34 fields (i.e. where we have $Y_{105w}$ imaging) this $zYJ$ selection selection criteria is:
\begin{eqnarray*} 
(z_{850lp}-Y_{105w})& > & 1.0\\
(z_{850lp}-Y_{105w})& > & 2.4\times (Y_{105w}-J_{125w})+0.4\\
(Y_{105w}-J_{125w}) & < & 1.0\\
\end{eqnarray*} 

The use of an alternative $Y$ filter ($Y_{098m}$) in the ERS field necessitates the use of a slightly different criteria:
\begin{eqnarray*} 
(z_{850lp}-Y_{098m})& > & 1.2\\
(z_{850lp}-Y_{098m})& > & 0.9\times (Y_{098m}-J_{125w})+0.7\\
(Y_{098m}-J_{125w}) & < & 2.0\\
\end{eqnarray*} 

It is important to note that such criteria biases against a hypothetical population of $z\approx 7$ galaxies which have extremely red UV spectral slopes (resulting in $J_{125w}-H_{160w}>1.0$ colours), which could be attributed to the presence of large amounts of dust attenuation. The potential bias of our selection criteria and a more general analysis of the UV properties of the candidates presented in this work is discussed in more detail in Wilkins et al. (2010b) where we conclude that the distribution of UV spectral slope indices is consistent with $\beta\approx-2.2$ (i.e. rather blue).

A second consideration to note is that while our selection criteria excludes all known intrinsic contaminants, it remains possible that there exists a hitherto unknown population of objects whose intrinsic photometry may be consistent with our selection criteria.

\subsubsection{Photometric Scatter}

While the selection criteria described in the preceding section is able to guard against contaminants whose intrinsic photometry matches that which is observed, photometric noise may scatter objects into our selection window, providing a significant source of contamination.

To asses the impact of photometric scatter when working at low signal-to-noise, we performed a simple simulation to generate predictions at fainter magnitudes based on the colour distribution of more luminous objects. Using the observed $bvizYJH$ photometry distribution of bright sources ($24.0<J_{125w}<26.0$ in the case of the HUDF) synthetic objects were inserted into the original images with fainter magnitudes. The number of inserted synthetic sources in a given magnitude interval was chosen to recreate the observed number of sources in the same interval.

\begin{figure}
\centering
\includegraphics[width=19pc]{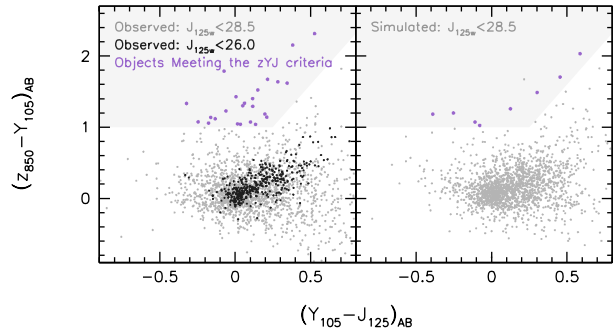}
\caption{Comparison of the observed (left panel) $(z_{850lp} - Y_{105w})$ -  $(Y_{105w} - J_{125w})$ colours of objects in the HUDF with that simulated by adding noise to observed distribution of bright galaxies (right panel). In both cases objects meeting our $zYJ$ criteria are indicated by the enlarged points.}
\label{fig:photo_s}
\end{figure}

Selecting objects in the HUDF with $J_{125w}<28.5$, suggests, from this simulation, that $\sim 14$ objects would scatter into our $zYJ$ selection window, surpassing the number of potential high-$z$ candidates. This is shown in Figure \ref{fig:photo_s} where, in the right panel, the simulated colours of $J_{125w}<28.5$ objects are shown and in the left the observed distribution (again $J_{125w}<28.5$)

To guard against this serious level of contamination we take advantage of the deep optical imaging that is available in the ACS $b_{435w}$, $v_{606w}$ and $i_{775w}$ bands. Star forming galaxies at $z\approx 7$ are not expected to contribute any flux in these bands as the lie blueward of the rest-frame location of the Lyman-$\alpha$ $1216{\rm \AA}$, thus we impose an additional $bvi$ {\em non-detection} criteria for the selection of our candidates. That is, we classify any object with a $>2\sigma$ detection in any of the $b_{435w}$, $v_{606w}$ and $i_{775w}$ filters as a lower-redshift contaminant. Imposing this selection criteria into our simulation we then find significantly fewer potenitial contaminants ($\sim 0.6$ in the HUDF). 

Making the $J_{125w}$ magnitude limit fainter by $0.5$ mag results in the contamination rate jumping significantly. This analysis highlights the necessity of having additional deep optical imaging to remove any such contamination. In total, over all fields, we expect a contamination rate of $\sim 2.4$ objects due to photometric scatter (or a $\sim 5\%$ contamination rate) (assuming a $J_{125w}$ limit corresponding to the average $6\sigma$ depth).

\subsubsection{Transients}

The large time delay between the acquisition of the ACS $z_{850lp}$ (taken between 2004 and 2006) and the WFC3 $YJH$ imaging leaves open the possibility of contamination by transient phenomena such as supernovae (SNe) and objects with significant apparent motions. Indeed, it appears both these phenomena are  contaminants within our fields.

Supernovae occurring within the WFC3 observation window could produce significant flux in the $YJH$ bands. This would naturally produce a large flux increase between the ACS $bviz$-bands (in which the transient would not contribute) and the NIR filters, mimicking the effect of a (potentially) very strong spectral break. The analysis of Bouwens et al. (2008) suggests a supernovae contamination rate of $0.012$ per arcmin$^{2}$ suggesting that we would expect to see $\sim0.75$ SN over all the fields.

Contamination from supernovae can, however, at the bright end be guarded against by comparing the current NIR imaging with pre-exist NIR imaging. The brightest preliminary candidate in the HUDF (which is not included in our final list of candidates), with $J_{125w}=25.6$ (flagged as zD0 in Bunker et al. 2010 and noted in Oesch et al. 2010, McLure et al. 2010, Finkelstein et al. 2010 and Yan et al. 2010) should be detected in previous NICMOS imaging of the field (which has a $2\sigma$ depth of $J_{110w}=26.7$) had it been a persistent object (i.e. not a transient). Its absence in this additional imaging is used as evidence that it is a contaminant and it is thus excluded in our analysis (as in Bunker et al. 2010).

\begin{figure}
\centering
\includegraphics[width=14pc]{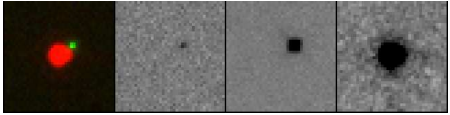}
\caption{$2\farcs.4\times2\farcs.4$ composite (B=$v$, G=$iz$, R=$YJ$) and $izY$ thumbnail images of a significant apparent motion object in the P12 field.}
\label{fig:pm_star}
\end{figure}

Phenomena with significant apparent motions are possible contaminants as the position of the object when the $z$-band imaging was acquired will be offset from that when the WFC3 NIR imaging was obtained. Photometry based on the location of the source in the $Y$-band image could then artificially suggest a break between the {\em WFC3} and {\em ACS} bands. Through visual inspection of the $z_{850lp}$ and $Y_{098m/105w}$ images for each candidate such objects can be guarded against. One such object was found in the P12 field (thumbnails of the RGB composite and $izY$ images is shown in Figure \ref{fig:pm_star}) and was removed from subsequent analysis.

\subsection{$z\approx 7$ Candidate Galaxies}

\begin{figure}
\centering
\includegraphics[width=18pc]{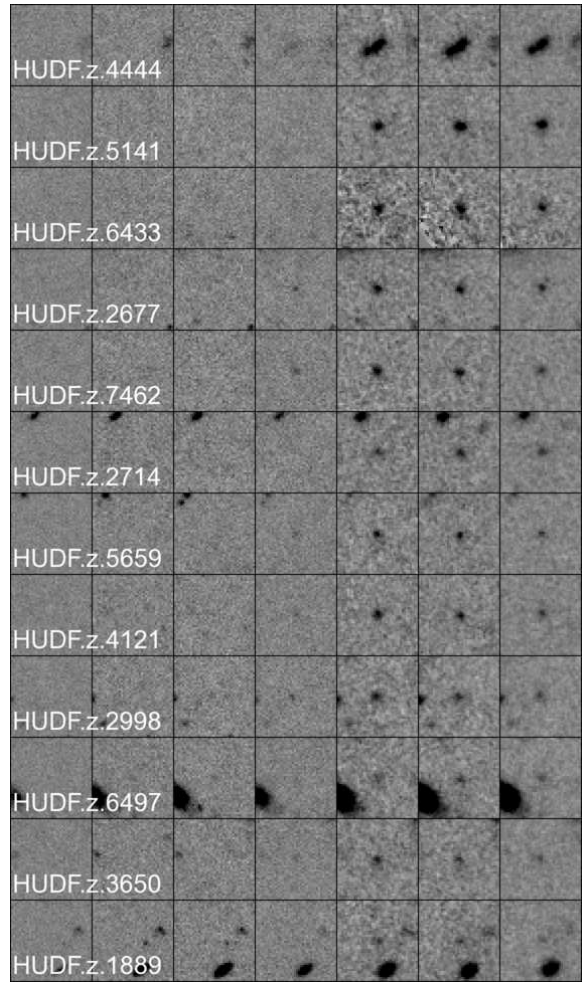}
\caption{$2\farcs4\times 2\farcs4$ $bvizYJH$ thumbnail images of objects meeting our selection criteria in the HUDF field.}
\label{fig:t_HUDF}
\end{figure}

\begin{figure}
\centering
\includegraphics[width=18pc]{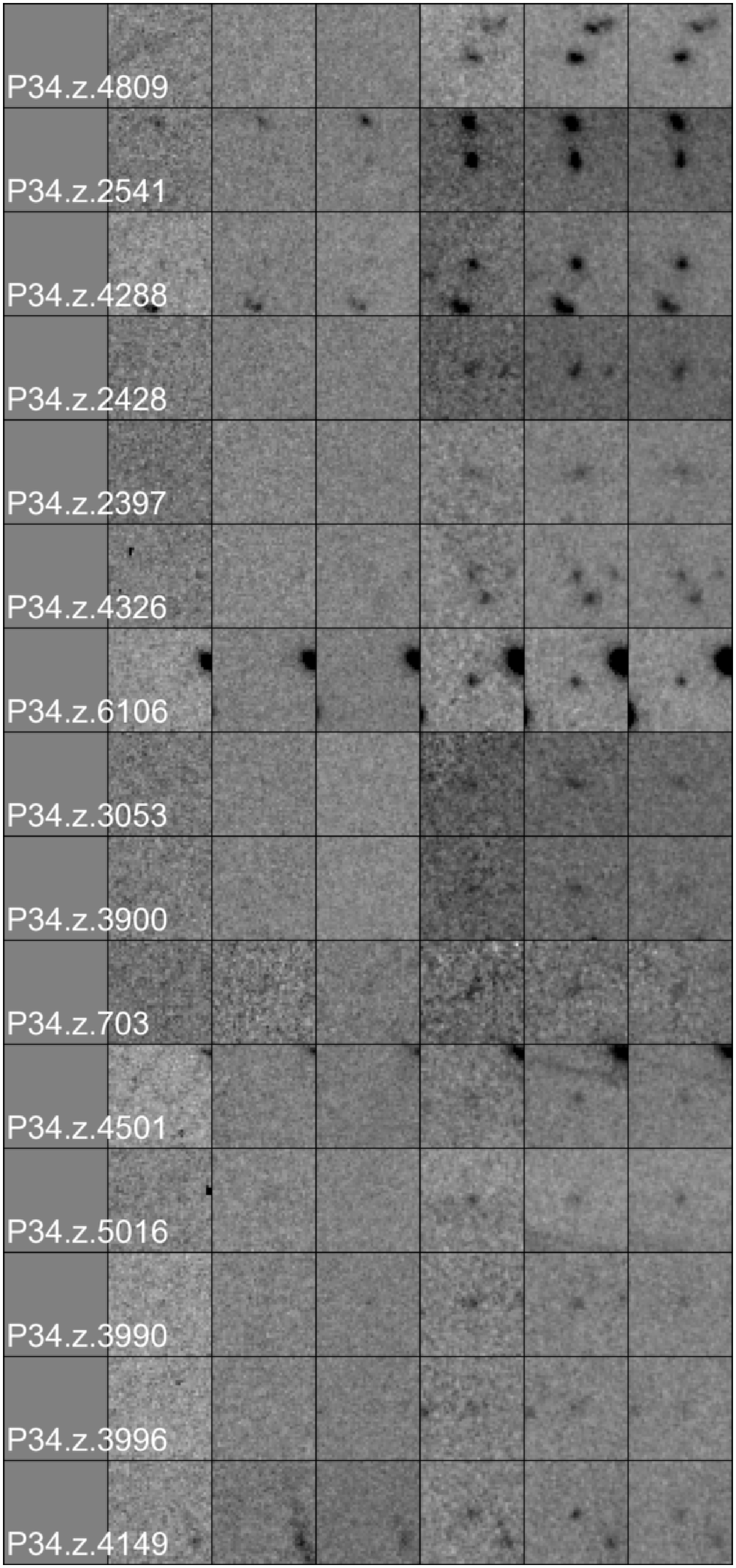}
\caption{$2\farcs4\times 2\farcs4$ $vizYJH$ thumbnail images of objects meeting our selection criteria in the P34 field.}
\label{fig:t_P34}
\end{figure}

\begin{figure}
\centering
\includegraphics[width=18pc]{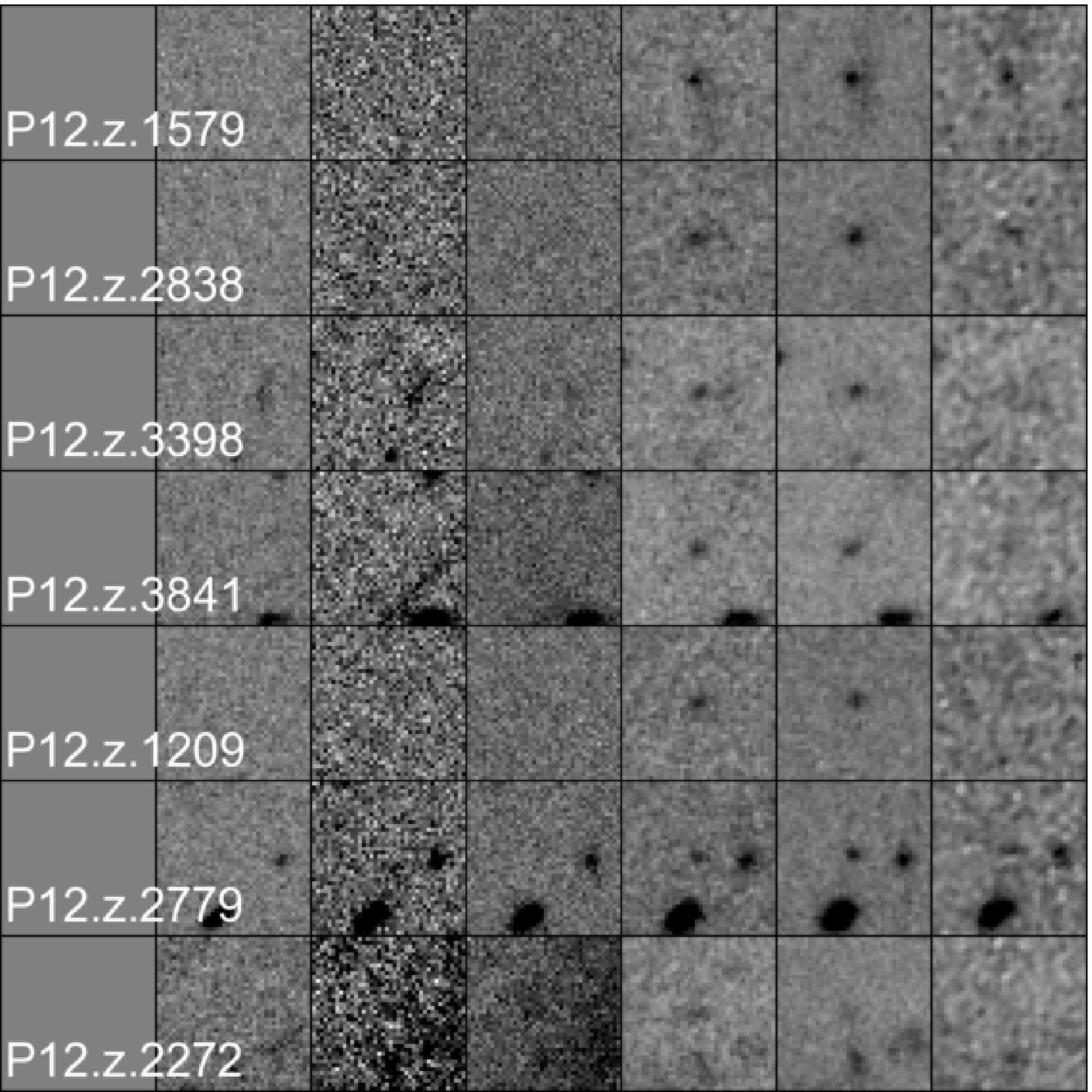}
\caption{$2\farcs4\times 2\farcs4$ $vizYJH$ thumbnail images of objects meeting our selection criteria in the P12 field.}
\label{fig:t_P12}
\end{figure}

\begin{figure}
\centering
\includegraphics[width=18pc]{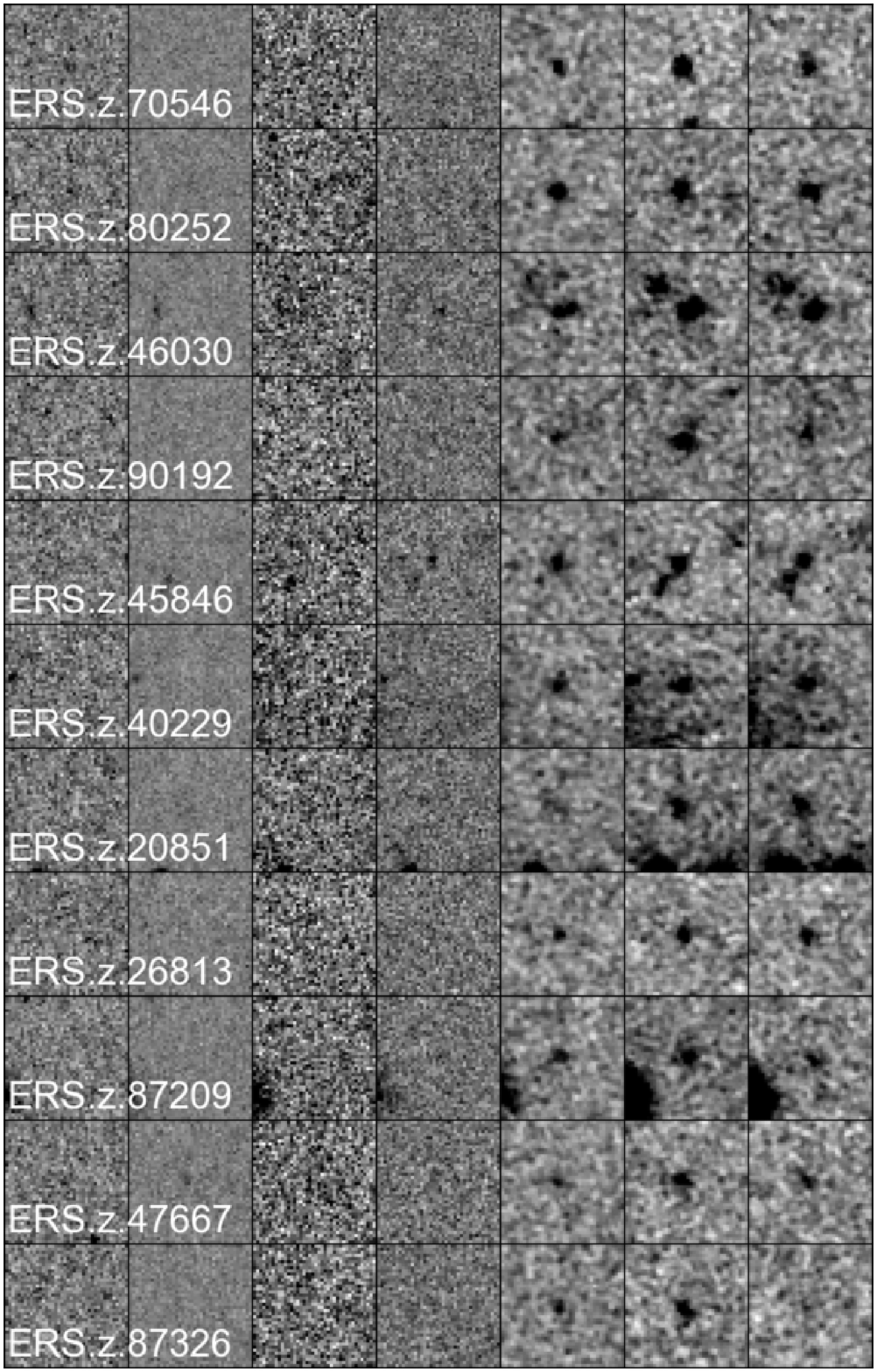}
\caption{$2\farcs4\times 2\farcs4$ $bvizYJH$ thumbnail images of objects meeting our selection criteria in the ERS field.}
\label{fig:t_ERS}
\end{figure}

Using the selection criteria described in the preceding subsection we compile a list of candidate $z\approx 7$ star forming galaxies in the HUDF, P34, P12 and ERS fields. The positions and photometry of these objects are listed in Table \ref{tab:objects}. Thumbnails of the $bvizYJH$ images of these candidates (where available) are presented in Figures \ref{fig:t_HUDF} - \ref{fig:t_ERS}. In total we find 44 (HUDF:11, P34:15, P12:7, ERS:11) candidates covering a range of apparent $J_{125w}$ magnitudes of $25.9-28.36$. A cross-matching of these candidates with those presented in other studies is undertaken in Section 4.

\begin{table*}
\begin{tabular}{lrrcccccc}
ID & RA Dec$^{a}$ & $z_{\rm AB}$ & $Y_{098m}/Y_{105w}$ & $J_{125w}$ & $H_{160w}$ & $(z-Y)_{AB}$ & $(Y-J)_{AB}$ \\
\hline\hline
HUDF.z.4444 & 03:32:42.553 -27:46:56.567 & $28.44 \pm 0.16$ & $26.71 \pm 0.03$ & $26.44 \pm 0.03$ & $26.41 \pm 0.02$ & $1.73$ & $0.27$ \\
HUDF.z.5141 & 03:32:38.804 -27:47:07.192 & $ > 30.01$ & $27.48 \pm 0.06$ & $26.9 \pm 0.04$ & $26.76 \pm 0.03$ & $ >2.52$ & $0.58$ \\
HUDF.z.6433 & 03:32:42.563 -27:47:31.442 & $29.18 \pm 0.31$ & $27.5 \pm 0.07$ & $27.1 \pm 0.05$ & $27.3 \pm 0.05$ & $1.68$ & $0.39$ \\
HUDF.z.2677 & 03:32:42.192 -27:46:27.858 & $28.66 \pm 0.2$ & $27.64 \pm 0.07$ & $27.7 \pm 0.08$ & $27.72 \pm 0.07$ & $1.02$ & $-0.06$ \\
HUDF.z.7462 & 03:32:36.772 -27:47:53.600 & $28.83 \pm 0.23$ & $27.64 \pm 0.07$ & $27.75 \pm 0.08$ & $27.82 \pm 0.08$ & $1.2$ & $-0.11$ \\
HUDF.z.2714 & 03:32:43.134 -27:46:28.474 & $ > 29.94$ & $28.15 \pm 0.12$ & $27.82 \pm 0.09$ & $27.73 \pm 0.07$ & $ >1.85$ & $0.33$ \\
HUDF.z.5659 & 03:32:36.378 -27:47:16.235 & $29.29 \pm 0.34$ & $28.15 \pm 0.12$ & $28.0 \pm 0.11$ & $28.25 \pm 0.12$ & $1.13$ & $0.15$ \\
HUDF.z.4121 & 03:32:37.443 -27:46:51.291 & $28.97 \pm 0.26$ & $27.83 \pm 0.09$ & $28.03 \pm 0.11$ & $27.97 \pm 0.09$ & $1.13$ & $-0.2$ \\
HUDF.z.2998 & 03:32:43.780 -27:46:33.706 & $28.94 \pm 0.25$ & $27.89 \pm 0.09$ & $28.05 \pm 0.11$ & $28.13 \pm 0.11$ & $1.05$ & $-0.15$ \\
HUDF.z.6497 & 03:32:36.463 -27:47:32.409 & $29.89 \pm 0.59$ & $28.36 \pm 0.14$ & $28.31 \pm 0.14$ & $28.18 \pm 0.11$ & $1.53$ & $0.06$ \\
HUDF.z.3650 & 03:32:40.565 -27:46:43.575 & $29.48 \pm 0.41$ & $28.09 \pm 0.11$ & $28.36 \pm 0.15$ & $28.42 \pm 0.14$ & $1.39$ & $-0.27$ \\
$[${\em HUDF.z.1889}$^{b}$ & 03:32:41.823 -27:46:11.275 & $29.71 \pm 0.5$ & $28.43 \pm 0.15$ & $28.44 \pm 0.16$ & $28.91 \pm 0.22$ & $1.29$ & $-0.01$$]$ \\
\hline
P34.z.4809 & 03:33:03.781 -27:51:20.475 & $ > 29.54$ & $26.97 \pm 0.06$ & $26.39 \pm 0.03$ & $26.29 \pm 0.03$ & $ >2.73$ & $0.59$ \\
P34.z.2541 & 03:33:09.640 -27:50:50.898 & $28.28 \pm 0.2$ & $26.47 \pm 0.04$ & $26.4 \pm 0.03$ & $26.39 \pm 0.03$ & $1.81$ & $0.07$ \\
P34.z.4288 & 03:33:09.879 -27:51:22.506 & $28.81 \pm 0.32$ & $26.93 \pm 0.06$ & $26.61 \pm 0.03$ & $26.54 \pm 0.04$ & $1.88$ & $0.32$ \\
P34.z.2428 & 03:33:09.763 -27:50:48.652 & $29.17 \pm 0.44$ & $27.64 \pm 0.11$ & $27.31 \pm 0.07$ & $27.02 \pm 0.06$ & $1.53$ & $0.33$ \\
P34.z.2397 & 03:33:02.182 -27:50:48.125 & $ > 29.63$ & $27.76 \pm 0.12$ & $27.38 \pm 0.07$ & $27.39 \pm 0.08$ & $ >1.94$ & $0.38$ \\
P34.z.4326 & 03:33:01.186 -27:51:13.395 & $28.72 \pm 0.29$ & $27.5 \pm 0.1$ & $27.41 \pm 0.07$ & $27.25 \pm 0.07$ & $1.22$ & $0.09$ \\
P34.z.6106 & 03:33:09.136 -27:51:55.548 & $29.2 \pm 0.45$ & $27.24 \pm 0.08$ & $27.46 \pm 0.07$ & $27.31 \pm 0.08$ & $1.96$ & $-0.22$ \\
P34.z.3053 & 03:33:11.166 -27:50:58.464 & $ > 29.63$ & $28.06 \pm 0.16$ & $27.67 \pm 0.09$ & $27.9 \pm 0.13$ & $ >1.64$ & $0.39$ \\
P34.z.3900 & 03:33:11.236 -27:51:12.537 & $ > 29.64$ & $27.9 \pm 0.14$ & $27.68 \pm 0.09$ & $27.9 \pm 0.13$ & $ >1.8$ & $0.22$ \\
P34.z.703 & 03:33:06.444 -27:50:11.449 & $29.21 \pm 0.46$ & $27.71 \pm 0.12$ & $27.92 \pm 0.11$ & $27.78 \pm 0.12$ & $1.5$ & $-0.21$ \\
P34.z.4501 & 03:33:07.287 -27:51:25.144 & $ > 29.51$ & $27.81 \pm 0.13$ & $27.95 \pm 0.12$ & $27.83 \pm 0.12$ & $ >1.89$ & $-0.14$ \\
P34.z.5016 & 03:33:03.301 -27:51:33.715 & $29.49 \pm 0.59$ & $27.67 \pm 0.11$ & $28.07 \pm 0.13$ & $27.69 \pm 0.11$ & $1.82$ & $-0.4$ \\
P34.z.3990 & 03:33:07.172 -27:51:16.280 & $29.43 \pm 0.56$ & $27.52 \pm 0.1$ & $28.09 \pm 0.13$ & $28.0 \pm 0.14$ & $1.91$ & $-0.58$ \\
P34.z.3996 & 03:33:07.086 -27:51:16.180 & $ > 29.59$ & $27.93 \pm 0.14$ & $28.15 \pm 0.14$ & $28.22 \pm 0.18$ & $ >1.77$ & $-0.22$ \\
P34.z.4149 & 03:33:05.395 -27:51:18.973 & $29.46 \pm 0.58$ & $27.93 \pm 0.15$ & $28.15 \pm 0.14$ & $27.89 \pm 0.13$ & $1.53$ & $-0.22$ \\
\hline
P12.z.1579 & 03:32:59.696 -27:40:35.243 & $28.98 \pm 0.39$ & $27.15 \pm 0.09$ & $26.85 \pm 0.05$ & $26.94 \pm 0.11$ & $1.83$ & $0.3$ \\
P12.z.2838 & 03:32:56.700 -27:41:08.205 & $29.02 \pm 0.41$ & $27.43 \pm 0.11$ & $27.1 \pm 0.06$ & $27.36 \pm 0.16$ & $1.59$ & $0.33$ \\
P12.z.3398 & 03:32:59.594 -27:41:20.916 & $ > 29.52$ & $27.69 \pm 0.14$ & $27.33 \pm 0.08$ & $27.58 \pm 0.2$ & $ >1.94$ & $0.36$ \\
P12.z.3841 & 03:33:02.433 -27:41:31.318 & $28.84 \pm 0.35$ & $27.67 \pm 0.14$ & $27.56 \pm 0.09$ & $27.68 \pm 0.22$ & $1.17$ & $0.11$ \\
P12.z.1209 & 03:32:58.503 -27:40:23.896 & $ > 29.63$ & $27.63 \pm 0.13$ & $27.73 \pm 0.11$ & $ > 28.3$ & $ >2.0$ & $-0.11$ \\
P12.z.2779 & 03:32:55.751 -27:41:07.030 & $ > 29.58$ & $28.0 \pm 0.19$ & $27.78 \pm 0.11$ & $27.32 \pm 0.16$ & $ >1.63$ & $0.22$ \\
P12.z.2272 & 03:33:03.606 -27:40:54.485 & $ > 29.13$ & $27.82 \pm 0.16$ & $27.98 \pm 0.14$ & $ > 28.3$ & $ >1.81$ & $-0.16$ \\
\hline
ERS.z.70546 & 03:32:27.908 -27:41:04.226 & $ > 28.43$ & $26.78 \pm 0.09$ & $25.97 \pm 0.03$ & $26.51 \pm 0.07$ & $ >1.68$ & $0.81$ \\
ERS.z.80252 & 03:32:12.924 -27:41:41.324 & $27.87 \pm 0.38$ & $26.11 \pm 0.05$ & $26.1 \pm 0.04$ & $26.15 \pm 0.05$ & $1.76$ & $0.01$ \\
ERS.z.46030 & 03:32:22.667 -27:43:00.700 & $27.93 \pm 0.4$ & $26.36 \pm 0.06$ & $26.11 \pm 0.04$ & $25.95 \pm 0.04$ & $1.57$ & $0.25$ \\
ERS.z.90192 & 03:32:24.094 -27:42:13.852 & $28.33 \pm 0.58$ & $26.96 \pm 0.11$ & $26.48 \pm 0.05$ & $26.73 \pm 0.08$ & $1.37$ & $0.48$ \\
ERS.z.45846 & 03:32:15.997 -27:43:01.446 & $27.85 \pm 0.38$ & $26.36 \pm 0.06$ & $26.48 \pm 0.05$ & $26.47 \pm 0.06$ & $1.49$ & $-0.13$ \\
ERS.z.40229 & 03:32:25.285 -27:43:24.251 & $ > 28.42$ & $26.6 \pm 0.08$ & $26.52 \pm 0.05$ & $26.28 \pm 0.05$ & $ >1.86$ & $0.08$ \\
ERS.z.20851 & 03:32:29.078 -27:44:30.687 & $28.42 \pm 0.63$ & $27.2 \pm 0.13$ & $26.72 \pm 0.06$ & $26.54 \pm 0.07$ & $1.21$ & $0.48$ \\
ERS.z.26813 & 03:32:22.935 -27:44:09.910 & $28.4 \pm 0.62$ & $27.17 \pm 0.13$ & $26.83 \pm 0.07$ & $26.71 \pm 0.08$ & $1.23$ & $0.34$ \\
ERS.z.87209 & 03:32:29.541 -27:42:04.492 & $ > 28.34$ & $26.93 \pm 0.1$ & $26.84 \pm 0.07$ & $27.04 \pm 0.11$ & $ >1.53$ & $0.09$ \\
ERS.z.47667 & 03:32:15.666 -27:42:53.562 & $ > 28.42$ & $27.12 \pm 0.12$ & $26.89 \pm 0.07$ & $27.08 \pm 0.11$ & $ >1.34$ & $0.23$ \\
ERS.z.87326 & 03:32:23.154 -27:42:04.654 & $ > 28.44$ & $27.21 \pm 0.13$ & $26.96 \pm 0.08$ & $27.76 \pm 0.21$ & $ >1.25$ & $0.25$ \\
\hline
\end{tabular}
\caption{z -band drop out candidate $z\approx7$ galaxies meeting our selection criteria in the HUDF, P34, P12 and ERS fields. Objects are divided by field and then ordered by apparent $J_{125w}$ magnitude. Photometry based on 0\farcs.6 diameter apertures with an aperture correction. Quoted limits correspond to $2\sigma$. $^{a}$ We introduce a small offset to the astrometry in the HST image headers of P34 of $\Delta RA=0\fs03$ to match the astrometry from GSC-2 and 2MASS. $^{b}$ This object has a $J_{125w}$ magnitude $0.1$ fainter than the adopted $7\sigma$ limit and was erroneously included in the initial draft of this work. It is omitted from all subsequent analysis.}
\label{tab:objects}
\end{table*}

\subsection{Comparison with Other Studies}

Both the WFC3 HUDF data (Bunker et al. 2010, Finkelstein et al. 2010, McLure et al. 2010, Oesch et al. 2010 and Yan et al. 2010) summarised in Table \ref{tab:others}) and some of the ERS pointings (Wilkins et al. 2010) have been previously analysed. In each case different reduction techniques, methods for measuring the photometry (including the choice of photometric zeropoints) and $z\approx 7$ selection criteria were used. A brief summary of these studies is given in Table \ref{tab:others} where information about their selection methodology and criteria is documented.

In the case of ERS dataset (partially analysed by Wilkins et al. 2010) we match 4 of the 6 candidates ($1$, $2$, $4$ and $6$) with the remaining two failing to make our $J_{125w}$ selection criteria (Wilkins et al. 2010 employed a $Y_{098m}$ criteria).

\begin{table*}
\begin{tabular}{lllll}
 & Study & Method & magnitude limit \\
\hline\hline
  -    & This Study & $z-Y$+$Y-J$ colour criteria + $bvi$ $<2\sigma$& $J_{125w}<28.5$ \\
B10$^{a}$ & Bunker et al. 2010 & $z-Y$+$Y-J$ colour criteria + $bvi$ $<2\sigma$ & $Y_{105w}<28.5$ \\
O10$^{a}$ & Oesch et al. 2010 & $z-Y$+$Y-J$ colour criteria + $bvi$ $<2\sigma$ & $S/N(Y_{105w}\land  J_{125w})>5\sigma$\\
M10$^{a}$ & McLure et al. 2010 & $bvizYJH$ photometric redshifts & $S/N(Y_{105w}\lor  J_{125w}\lor H_{160w})>5\sigma$ \\
F10 & Finkelstein et al. 2010 & $bvizYJH$ photometric redshifts & $S/N(J_{125w}\land H_{160w})>3.5\sigma$\\
Y10 & Yan et al. 2010 & $z-Y$ colour criteria  $bvi$ $<2\sigma$ & $S/N(Y_{105w})>3.0$\\
\end{tabular}
\caption{Summary of other studies that analysed the HUDF field. $^{a}$ Employed different zeropoints.}
\label{tab:others}
\end{table*}

\begin{table*}
\begin{tabular}{lllllll}
ID & B10 & M10 & O10 & F10 & Y10 & Others\\
\hline\hline
HUDF.z.4444 & zD1 & 688z (6.7) & UDFz-42566566 & 1441 (6.84) & z7-A032 \\
HUDF.z.5141 & zD2 & 835z (7.2) & UDFz-38807073 & 1768 (7.22) & z7-A025 \\
HUDF.z.6433 & zD3 & 1144z (6.8) & UDFz-42577314 & 2432 (6.86) & z7-A008 \\
HUDF.z.2677 & - & 1464 (6.3) & - & 649 (6.45) & - \\
HUDF.z.7462 & - & 1911z (6.4) & UDFz-36777536 & - & - \\
HUDF.z.2714 & zD5 & 1678z (7.05) & UDFz-43146285 & 669 (7.3) & z7-A060 \\
HUDF.z.5659 & zD6 & 1958z (6.5) & UDFz-36387163 & 2032 (6.4) & z7-A016 \\
HUDF.z.4121 & - & 1880z (6.5) & UDFz-37446513 & 1289 (6.34) & z7-A040 \\
HUDF.z.2998 & - & 1855 (6.4) & - & 803 (6.4) & - \\
HUDF.z.6497 & - & 2003 (6.3) & - & 2465 (6.56) & - \\
HUDF.z.3650 & zD8 & 2206z (6.5) & UDFz-40566437 & 1072 (6.49) & z7-A047 \\
\end{tabular}
\caption{A cross-matched list of our candidates and those reported by the studies listed in Table \ref{tab:others} and any additional studies. Numbers in parentheses in the McLure et al. (2010) (M10) and Finkelstein et al. (2010) (F10) columns denote the photometric redshifts recovered by these studies.}
\label{tab:ourcomp}
\end{table*}

Cross-matching our candidates with those presented in the studies listed in Table \ref{tab:others} indicates that all our HUDF candidates exist in at least two of the other studies. A cross matched list of our candidates and the corresponding identifications from the other studies are shown in Table \ref{tab:ourcomp}.

Due to differences in the selction technique (and criteria), there are several candidate objects included in the studies listed in Table \ref{tab:others} which fail to enter our selection. These are listed and cross-matched in Table \ref{tab:theircomp}. The McLure et al. (2010) and Finkelstein et al. (2010) analysis used a photometric redshift selection technique; as such their catalogues may include objects at too high, or too low redshift to meet our selection criteria, depending on their choice of redshift range. In Tables \ref{tab:ourcomp} and \ref{tab:theircomp} only objects with photmetric redshifts in the range $6.3<z<7.8$ are included. This range is motivated by a series of simulations discussed in more depth in Section \ref{sec:red}.

A significant fraction of the objects included in Table \ref{tab:theircomp} are excluded from our selection on the basis of our conservative $J_{125w}$ limit. For example two objects (zD10 and zD7) in Bunker et al. (2010) (who adopt a $Y_{105w}$ magnitude limit criteria) are excluded on this basis ($5/8$ remaining objects from Oesch et al. 2010 are excluded for this reason, as are $4/11$ of the remaining Yan et al. 2010 objects). 

The remainder of the excluded objects fail to meet our $(z_{850lp}-Y_{105w}):(Y_{105w}-J_{125w})$ colour selection criteria (i.e. none are excluded on the basis of an optical detection). Bunker et al. (2010) initially employed a $z_{850lp}-Y_{105w}$ cut before considering the $Y_{105w}-J_{125w}$ colour to exclude very-red potential low-redshift and stellar interlopers. As a result several objects (zD11, zD9, zD4) which make our $J_{125w}$ selection criteria fail to make our colour selection criteria. On inspection of the photometry of these objects they are revealed to all lie very close to our selection window suggesting they are potential high-redshift galaxies and not contaminants. 

\begin{figure}
\centering
\includegraphics[width=18pc]{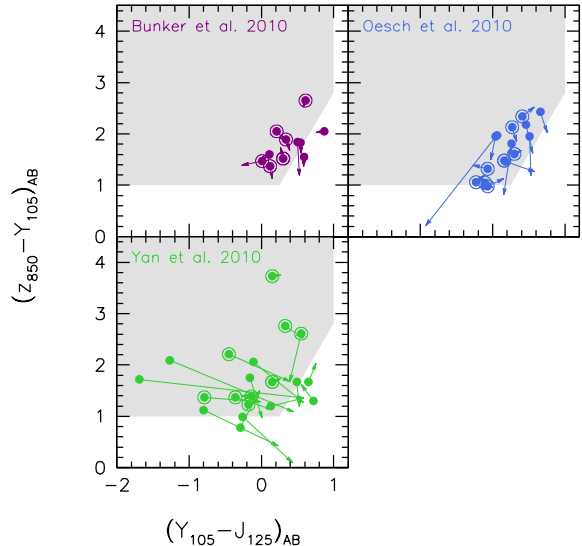}
\caption{Comparison of the $z_{850lp}-Y_{105w}$ and $Y_{105w}-J_{125w}$ colour for candidates reported by other studies from both the original photometry (i.e. from the other studies) and our photometry. In each case the original photometry is denoted by the position of the circle. The associated arrows point to the position in colour-colour space where the object would lie according to our photometry. Objects included as candidates in this study are circled.}
\label{fig:cccc}
\end{figure}

The exlcusion of these candidates can be seen more clearly in Figure \ref{fig:cccc}. In this figure the points denote the location (in the $z_{850lp}-Y_{105w}$ - $Y_{105w}-J_{125w}$ colour plane) of candidates from the Oesch et al. (2010), Bunker et al. (2010) and Yan et al. (2010) studies and the corresponding arrows point to the location based on our photometry. Those objects meeting our selection criteria {\em based on our photometry} are denoted with a {\em halo}. The figure shows that candidates failing to make our selection criteria generally lie close to the selection window. Figure \ref{fig:cccc} also highlights that there exist sometimes significant differences between the photometry of the various studies. To some extent this is due to differences in the assumed photometric zeropoints (i.e. no attempt was made to {\em harmonise} these) with the remainder of the deviations due to differences in the reduction process and choice of photometry methodology. 

Comparing our candidate list with those of McLure et al. (2010) and Finkelstein et al. (2010) is slightly more complicated as they employ photometric redshift techniques and present candidates encompassing a larger redshift range. Of their candidates lying within our predicted redshift selection window ($6.3<z<7.8$ , see Section \ref{sec:red}) several have $J_{125w}$ magnitudes fainter than our imposed limit. The remaining objects have $(z_{850lp}-Y_{105w}):(Y_{105w}-J_{125w})$ colours which are typically only marginally inconsistent with our selection criteria.

We have imposed a colour selection criteria and S/N limit where contamination from photometric scatter is minimal. Other groups have presented objects outside our colour selection, or at lower S/N which may plausibly be high-redshift, but have a greater possibility of being contaminants. The omission of these candidates from our selection, however, does not impact any of the conclusions drawn from the observed luminosity function (see Section \ref{sec:lf}). As described in the following sections, the luminosity function is determined by considering the effective volume of the survey, which is sensitive to the choice of selection criteria. Thus, our conservative selection criteria should recover the true luminosity function of $z\approx 7$ star forming galaxies.

\begin{table*}
\begin{tabular}{llllll}
B10 & M10 & O10 & F10 & Y10 & \\
\hline\hline
 zD4$\dagger$ & 1092z$\dagger$ (6.85) & UDFz-39557176$\dagger$ & 2056 (7.27) & z7-A014$\dagger$ & B \\
 zD7 & 1107z (7.6) & UDFz-44716442$\dagger$ & 1110 (6.67) & z7-A044 & AB \\
 zD9$\dagger$ & 1574z (7.2) & UDFz-37228061$\dagger$ & 3053 (7.4) & z7-A003 & B \\
 zD10$\dagger$ & 2502z (7.1) & UDFz-39736214 & 515 (6.88) & z7-A057 & A \\
 zD11 & - & - & 335 (6.86) & z7-A053 & B \\
 zD4$\dagger$ & 1092z$\dagger$ (6.85) & UDFz-39557176$\dagger$ & 2056 (7.27) & z7-A014$\dagger$ & B \\
 zD7 & 1107z (7.6) & UDFz-44716442$\dagger$ & 1110 (6.67) & z7-A044 & AB \\
 zD9$\dagger$ & 1574z (7.2) & UDFz-37228061$\dagger$ & 3053 (7.4) & z7-A003 & B \\
 zD10$\dagger$ & 2502z (7.1) & UDFz-39736214 & 515 (6.88) & z7-A057 & A \\
 zD11 & - & - & 335 (6.86) & z7-A053 & B \\
 - & 934i (6.2) & - & 769 (6.4) & - & B \\
 - & 2514 (6.3) & - & - & - & B \\
 - & 1864 (6.4) & - & - & - & B \\
 - & 1915z$\dagger$ (6.4) & UDFz-39586565 & 1445 (6.36) & z7-A033$\dagger$ & AB \\
 - & 2195 (6.45) & - & - & - & AB \\
 - & 1064$\dagger$ (6.65) & - & 1566$\dagger$ (6.54) & - & B \\
 - & 2794$\dagger$ (6.75) & - & - & - & B \\
 - & 2395 (6.8) & - & 1112 (6.34) & - & B \\
 - & 2560z (6.9) & UDFz-37807405$\dagger$ & 2644 (6.36) & - & AB \\
 - & 2826 (6.9) & - & - & - & B \\
 - & 2066z (7.2) & UDFz-41057156$\dagger$ & 2013 (6.79) & z7-A017$\dagger$ & B \\
 - & 2888 (7.35) & - & 1110 (6.67) & z7-A046 & B \\
 - & 2940 (7.4) & - & - & z7-A065 & B \\
 - & 2079y (7.5) & - & 213 (8.05) & - & B \\
 - & 2487 (7.8) & - & - & - & B \\
 - & - & UDFz-38537519 & - & - & AB \\
 - & - & - & 1818 (6.56) & - & A \\
 - & - & - & 567 (6.84) & - & A \\
 - & - & - & 457 (7.03) & - & A \\
 - & - & - & - & z7-A056 & B \\
 - & - & - & - & z7-A062 & AB \\
\end{tabular}
\caption{List of candidates from the studies outlined in Table \ref{tab:others} that fail to make it into our selection together with the reason why (final column). A - fails to meet our $J_{125w}$-band magnitude cut. B - fails to meet our colour selection. $\dagger$ would have met our selection criteria if photometry from original study was adopted.}
\label{tab:theircomp}
\end{table*}

\section{Candidate Properties}

\subsection{Surface Densities}

\begin{figure}
\centering
\includegraphics[width=20pc]{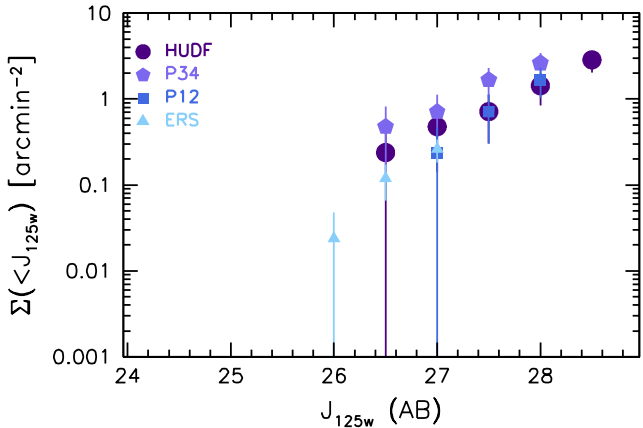}
\caption{The observed cumulative surface density (i.e. the number of objects brighter than the limiting $J_{125w}$ of $z\approx 7$ candidate galaxies for each of our fields. Note, because of the different $Y$ filters employed by the HUDF/P34/P12 and ERS fields the surface densities are not strictly comparable due to the slightly different redshift selection.}
\label{fig:sd}
\end{figure}

One useful diagnostic of the $z$-drop population probed in each field are the observed surface densities as a function of limiting $J_{125w}$-magnitude (as shown in figure \ref{fig:sd} for each field). This gives a direct indication of any field to field variance due to cosmic variance, that is whether one field is probing an area of higher-density than the average - a potential problem for narrow pencil-beam surveys. Further, knowledge of surface density is useful when considering the potential of high-redshift $z$-drop galaxies as contaminants in, for example, the search for very low-mass stars (e.g. Delorme et al. 2010).

The observed surface densities are consistent between fields (i.e. within the uncertainties) with the exception that the P34 surface dnesities exceed those of the ERS. However, this is unsurprising as the surface density of $z$-drops in the ERS is not directly comparable to that the other fields due to the use of the $Y_{098m}$ filter (over the $Y_{105w}$ used in the HUDF/P34/P12 fields). The agreement between the 3 deeper fields (HUDF/P34/P12), which are seperated by $\approx 2\,{\rm Mpc}$ suggests we are not probing a compact cluster which would affect our estimates of the rest-frame UV luminosity function and thus star formation rate density and ionising photon luminosity density.

\subsection{Completeness and Redshift Distributions}\label{sec:red}

\begin{figure}
\centering
\includegraphics[width=9pc]{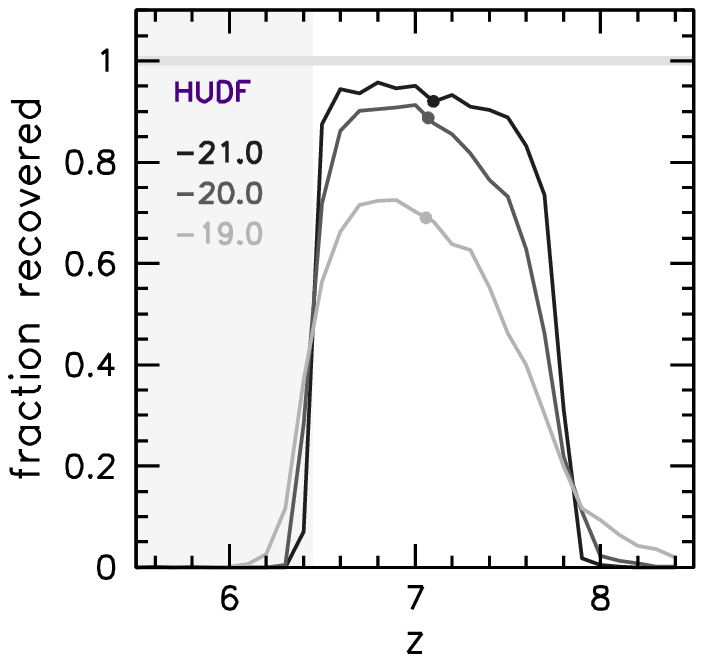}
\includegraphics[width=9pc]{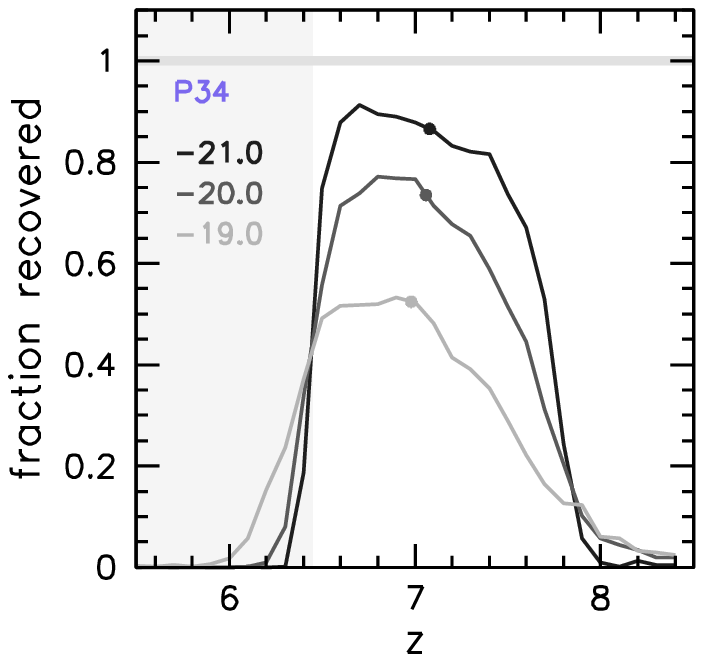}
\includegraphics[width=9pc]{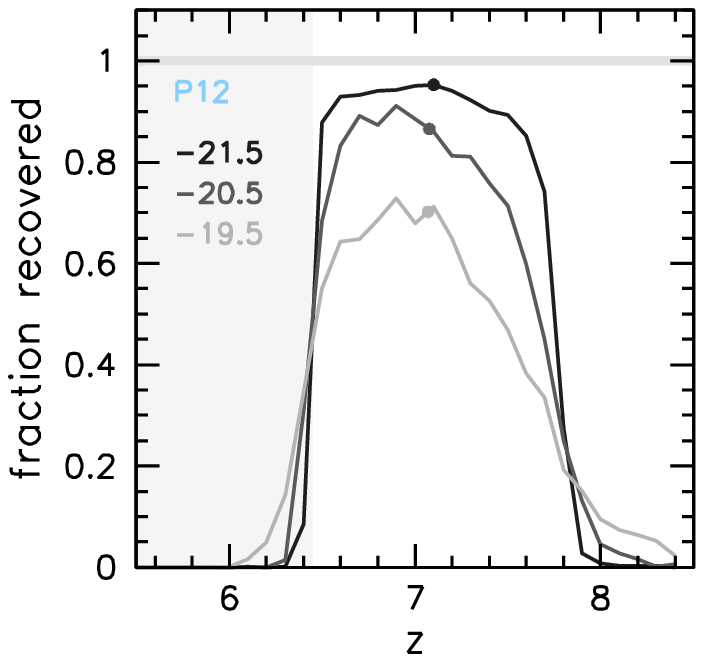}
\includegraphics[width=9pc]{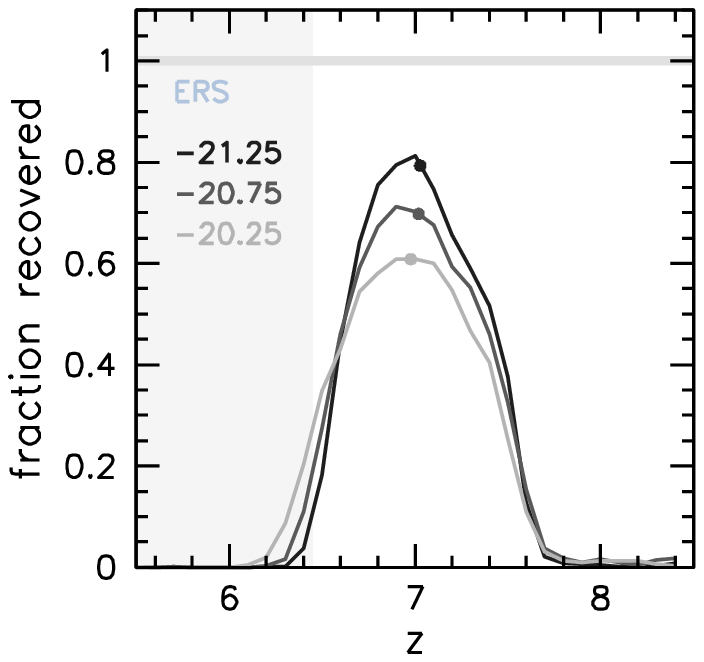}
\caption{The probability of recovering a galaxy in our simulation for each field as a function of redshift for several different absolute rest-frame M$_{1600}$ magnitudes. The mean redshift is in each case denoted by a dot.}
\label{fig:complete}
\end{figure}

A useful description of our $z$-drop population is the theoretical redshift distribution we would expect to probe, that is, it is useful to ascertain the probability of recovering a high-$z$ galaxy as a function of redshift and rest-frame UV luminosity - $p(M_{1600},z)$.

To quantify this probability a series of simulations is conducted for each field. These simulations are performed by inserting sources with properties similar to those expected for high-$z$ galaxies (i.e. compact sources with half-light radii $\sim 0.1"$) into the original images and running our selection process to determine the probability that a source would be recovered (and included in our selection). 

A further important parameter in the simulation is the choice of the UV continuum properties. This is parameterised throught the UV spectral slope index $\beta$ (where $f_{\lambda}\propto \beta^{\lambda}$ over the rest-frame UV continuum). Due to our selection criteria, objects with different values of $\beta$ enter the selection window at slightly different redshifts which would result in a different $p(M_{1600},z)$. However for objects with $-3.0 < \beta < 0.0$ the redshift at which objects enter the selection window is fairly uniform and the resulting values of $p(M_{1600},z)$ are very similar. Objects with $\beta > 0.0$ (suggestive of large dust attenuations) are however excluded as part of our strategy to eliminate low redshift contaminants.  However previous studies (e.g. Bouwens et al. 2009) have found that the UV slopes of high-redshift objects are typically blue with $\beta \approx-2.2$, corresponding to a star forming population which is essentially dust-free (under common assumptions regarding the previous star formation history, metallicity and initial mass function). For our candidates we find a mean $(J_{125w}-H_{160w})$ colour of $~0.0$ corresponding to $\beta=-2.0$ and find few objects with $(J_{125w}-H_{160w})>0.25$ (i.e. $\beta>-1.0$). In our simulations we assume that the UV spectral slopes of objects are distributed as a gaussian with $\langle\beta\rangle=-2.2$ and $\sigma=0.5$.  

The results of these simulations are shown in Figure \ref{fig:complete}. In the HUDF and flanking fields galaxies with large intrinsic luminosities ($M_{1600}<-20.0$) $p(M_{1600},z)$ remains fairly constant over $6.3<z<7.8$. At lower luminosities the probability of recovering a galaxy quickly decreases at the high-$z$ end of this range due to increasing luminosity distance and Lyman-alpha forest blanketting of the $Y_{105w}$-filter with redshift. The ERS field, by virtue of the narrower $Y_{098m}$, has a slightly more restricted redshift probability function than the other fields. Irrespective of the field, though with a slight bias to a higher-redshift for intrinsically more luminous galaxies, the expected mean redshift of $z_{850lp}$-drop high-$z$ star forming galaxies is $z\approx 7.0$.

\subsection{Luminosity Function}\label{sec:lf}

The probability of observing a galaxy as a function of absolute rest-frame UV magnitude and redshift, $p(M_{1600},z)$, described in the preceding section, can be used to generate a prediction of the effective volume surveyed as a function of absolute M$_{1600}$ magnitude which can then be used to estimate the piecewise luminosity function (LF). The relationship between the $p(M,z)$ and  effective volume $V_{\rm eff}$ is given by
\begin{equation} 
V_{\rm eff}=\int p(M,z)\frac{{\rm d}V}{{\rm d}z}{\rm d}z,
\end{equation} 
as described in (e.g. Steidel et al. 1999, Stanway et al. 2003). 

To estimate the LF we assume that all galaxies lie at the centre of the computed redshift distribution and determine the absolute magnitude. At $z=7.0$ the effective wavelength of the $J_{125w}$ filter corresponds to a rest-frame value of $1563{\rm \AA}$. For ease of comparison with other studies we $k$-correct the observed $J_{125w}$ magnitudes to $1600{\rm \AA}$ by using the observed $(J_{125w}-H_{160w})$ colour (though it is worth noting that this is a small correction given that  $(J_{125w}-H_{160w})$ is observed to be approximately $0.0$.

\subsubsection{Comparison With Other Studies}

\begin{figure}
\centering
\includegraphics[width=20pc]{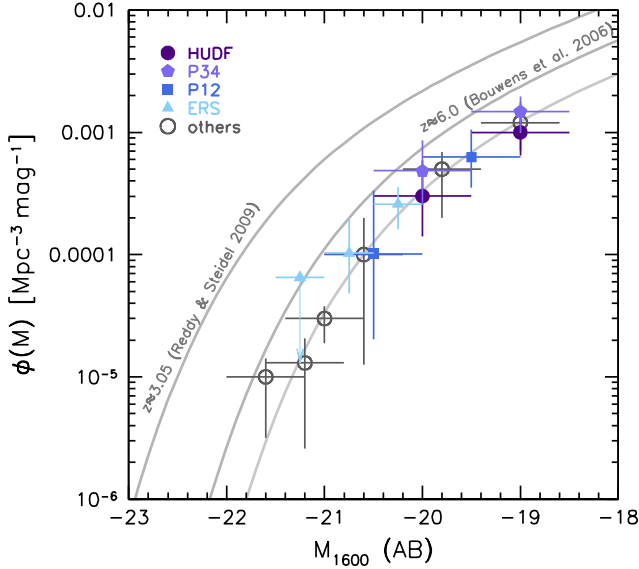}
\caption{The $z\approx 7$ rest-frame UV ($1600{\rm\AA}$) LF derived from each of the WFC3 fields (blue/purple points) together with contemporary and lower-redshift comparisons. The solid blue line denotes the best fit Schechter function assuming $\alpha=-1.7$. Solid grey lines denote the luminosity function at $\langle z\rangle=2.3$ (Reddy \& Steidel 2009) and $\langle z\rangle=6.0$ (Bouwens et al.). The grey circles are the estimates of Oesch et al. (2010) from their HUDF sample and from the wide area sample Ouchi et al. (2009). }
\label{fig:lf}
\end{figure}

The stepwise LF estimates for each field are mutually consistent within their uncertainties. Further, estimates from HUDF appear consistent with those of other authors (Oesch et al. 2010 and McLure et al. 2010, with only Oesch et al. 2010 shown in Figure \ref{fig:lf} for clarity) while the ERS constraints are consistent with the brighter sample of Ouchi et al. (2009) obtained using Subaru/Suprime-Cam.  

In order to explore evolution in the UV luminosity function, we compare our results with those at lower-redshift using the Schechter (1976) function fits to the $\langle z\rangle=3.05$ (Reddy \& Steidel 2009) and $\langle z\rangle=6.0$ (Bouwens et al. 2006) luminosity functions (included in Figure \ref{fig:lf}). The position of these LFs with respect to our estimates indicates a clear decline in the UV LF from $z=3\to 7$, together with significant evidence of evolution from $z=6\to 7$ (given that our stepwise LF estimates all lie below that of the Bouwens et al. 2006 $\langle z\rangle=6.0$ LF). 

\subsubsection{Schechter Function Parameterisation}

\begin{figure}
\centering
\includegraphics[width=20pc]{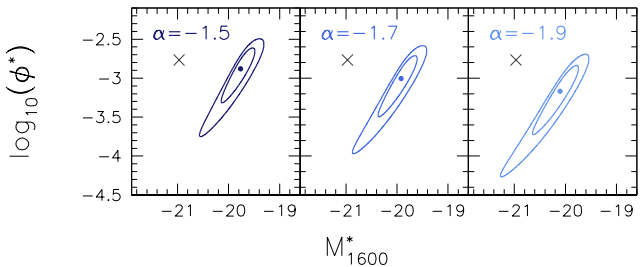}
\caption{$68\%$ and $95\%$ confidence contours for $M^{*}_{1600}$ and $\phi_{*}$ for each of the three assumed values of $\alpha$. The crosses mark the best-fit parameters found at $z\approx 3.05$ by Reddy \& Steidel (2009).}
\label{fig:conf_inter}
\end{figure}

To further explore the implications of the observed $z\approx 7$ luminosity function we fit the stepwise LF using the Schechter (1976) functional form of the luminosity function,
\begin{equation} 
\phi(L)=\frac{\phi^{*}}{L^{*}}\left(\frac{L}{L^{*}}\right)^{\alpha}exp\left( -L/L^{*}\right)
\end{equation} 
Unfortunately our stepwise LF estimates fail to extend sufficiently at lower luminosities to constrain the faint end slope of Schechter function. With this in mind we determine the remaining parameters, $M_{1600}^{*}$ and $\phi^{*}$, assuming a selection of different faint-end slope values consistent with lower redshift determinations $\alpha$ ($\in\{1.5,1.7,1.9\}$). The best fit values of $M_{1600}^{*}$ and $\phi^{*}$ assuming each of these values of $\alpha$ are listed in Table \ref{tab:uvld} and the $68\%$ and $95\%$ confidence intervals are shown in figure \ref{fig:conf_inter}.

\begin{table*}
\begin{tabular}{ccccccc}
& $M_{1600}^{*}$ & $\phi^{*}$ & $\alpha$ & \multicolumn{3}{|c|}{$\rho_{1600}\,[10^{25}\,{\rm erg\, s^{-1}\, Mpc^{-3}\, Hz^{-1}}]$ $(\dot{\rho_{*}}\,[{\rm M_{\odot}\,yr^{-1}\,Mpc^{-3}}])$} \\
& & & & $M_{1600}<-18.5$ $({\rm SFR}>1.5\,{\rm M_{\odot}\,yr^{-1}})$ & $<-13.0$ $(>0.01\,{\rm M_{\odot}\,yr^{-1}})$ & $<-8.0$ $(>0.0001\,{\rm M_{\odot}\,yr^{-1}})$ \\
\hline\hline
& -19.8 & 0.00126 & -1.5 & 3.6 ( 0.005 ) & 7.8 ( 0.010 ) & 8.2 ( 0.011 ) \\
& -19.9 & 0.00106 & -1.7 & 3.6 ( 0.005 ) & 10.6 ( 0.014 ) & 12.7 ( 0.016 ) \\
& -20.1 & 0.00072 & -1.9 & 3.6 ( 0.005 ) & 14.9 ( 0.019 ) & 25.1 ( 0.033 ) \\
\end{tabular}
\caption{The best fit values of $M^{*}_{1600}$ and $\phi_{*}$ assuming fixed $\alpha\in\{1.5,1.7,1.9\}$ together with the UV luminosity and star formation rate densities determined by integrating down to various depths.}
\label{tab:uvld}
\end{table*}

\subsubsection{UV Luminosity and Star Formation Rate Density}

\begin{figure}
\centering
\includegraphics[width=20pc]{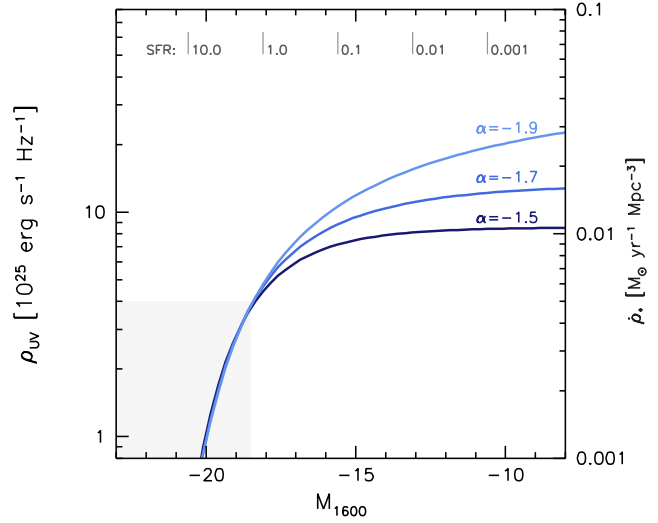}
\caption{The UV luminosity density (left axis) and star formation rate density (right axis) as a function of the $M_{1600}$ magnitude down to which the luminosity function is integrated for the best fit luminosity functions assuming $\alpha\in\{1.5,1.7,1.9\}$. The shaded grey box denotes the observed region with the remainder inferred from extrapolation of the LF.}
\label{fig:sfrd_z}
\end{figure}

An important quantity derived from the luminosity function is the ultraviolet luminosity density $\rho_{1600}=\int\,L\times\phi_{1600}(L)\,{\rm d}L$, from which the ionising photon luminosity and the star formation rate density can be inferred under a set of assumptions. Figure \ref{fig:sfrd_z} shows the UV luminosity density as a function of the magnitude down to which the luminosity function is integrated for each of the best-fit luminosity functions with fixed $\alpha\in\{1.5,1.7,1.9\}$. Table \ref{tab:uvld} includes the luminosity density for each of these LFs for $M_{1600}<-18.5$ (i.e. the {\em observed} luminosity density), $M_{1600}<-13.0$ and $M_{1600}<-8.0$.

Ultraviolet emission from star forming galaxies is dominated by massive ($m>5\,M_{\odot}$) short-lived stars. The short main-sequence lifetimes of these stars implies their presence, and thus UV emission, is a good tracer of the amount of ongoing star formation. The star formation rate of a galaxy is often related to the UV luminosity through a linear equation, i.e.
\begin{equation}
SFR=B_{\rm UV}\times 10^{-28}\,L_{\nu}\,\,[{\rm erg\,s^{-1}\,Hz^{-1}}],
\end{equation}
where the parameter $B_{\rm UV}$ is a calibration factor dependent on the previous star formation history, IMF and metallicity. 

Popular choices for this calibration parameter include $B_{\rm UV}=1.4$ (Kennicutt 1998) and $B_{\rm UV}=1.25$ (Madau, Pozetti \& Dickinson 1998). Using the {\sc Starburst99} (Leitherer et al. 1999) population synthesis model assuming solar metallicity, a Salpeter IMF and $100\,{\rm Myr}$ of previous star formation we obtain a value of $B_{\rm UV}=1.31$, similar to both the Kennicutt (1998) and Madau, Pozetti \& Dickinson (1998) calibrations. Applying this calibration factor we convert the UV luminosity density assuming the three luminosity functions obtained in the previous section to determine the star formation rate density. The relationship between star formation rate density and the magnitude to which the LF is integrated is shown in Figure \ref{fig:sfrd_z} (on the right axis). Table \ref{tab:uvld} also includes the star formation rate densities (where ${\rm SFR>}\in\{1.5,0.01,0.0001\}\,{\rm M_{\odot}\,yr^{-1}}$) for each of three best-fit luminosity function.

\begin{figure}
\centering
\includegraphics[width=20pc]{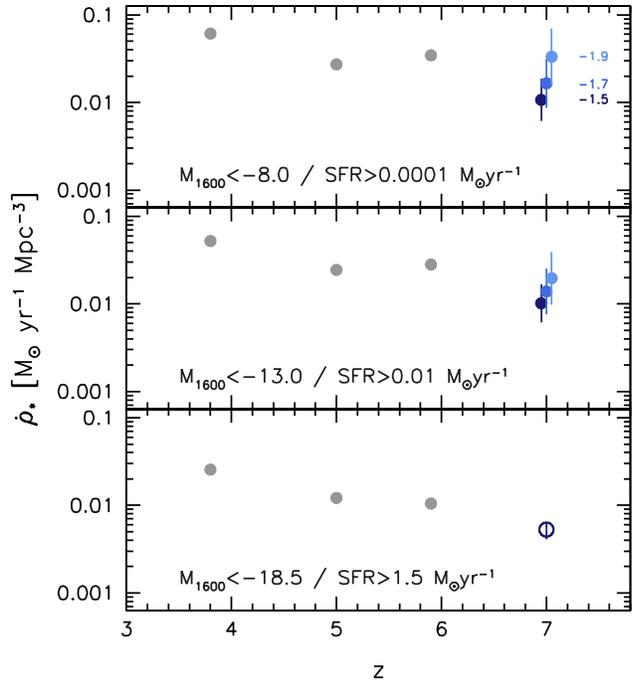}
\caption{The evolution of the very high-redshift ($z>3$) star formation rate density assuming various values of the limiting magnitude/star formation rate (${\rm SFR>}\in\{1.5,0.01,0.0001\}\,{\rm M_{\odot}\,yr^{-1}}$). $z\in\{3.8, 5.0, 5.9\}$ star formation rate densities are inferred from the Bouwens et al. (2007) UV luminosity functions. The $z\approx 7$ points are those inferred from the best-fit Schechter functions listed in Table \ref{tab:uvld}. For ${\rm SFR>}1.5\,{\rm M_{\odot}\,yr^{-1}}$ the star formation rate density is the same irrespective of the choice of faint-end slope.}
\label{fig:sfh}
\end{figure}

Using estimates of the UV luminosity function at lower redshifts ($z\in\{3.8, 5.0, 5.9\}$, from Bouwens et al. 2007) we can determine the dust-uncorrected high-redshift cosmic star formation rate density history. This is shown in Figure \ref{fig:sfh} again for 3 different $M_{1600}$ integration limits. Irrespective of the integration limit there is a clear decline in the star formation rate density $z=4\to 7$, assuming $\alpha<-1.7$. For steeper faint-end slopes the trend becomes somewhat less pronounced. Estimates of the dust attenuation from the distribution of UV spectral slopes (e.g. Bouwens et al. 2009, Bouwens et al. 2010b, Wilkins et al. {\em in prep}) suggest decreasing levels of attenuation at higher-redshift. Such a trend would further accentuate the declining trend in the star formation rate density over $z=4\to 7$.

\subsubsection{Reionisation}

As a result of the decreasing ambient temperature free electrons and protons recombined to form neutral hydrogen at $z\approx 1100$. Subsequently, the neutral hydrogen was reionised by a source(s) producing a sufficient number of ionising photons (i.e. with $\lambda<912{\rm\AA}$ for hydrogen). The identity of these sources, and the exact epoch in which transition took place, remain some of the key outstanding questions in astrophysics. One potential source of the ionising photons is the nascent $z>7$ star forming galaxy population, where the ionising photon flux comes from a very-massive short-lived stellar population. 

The number of ionising photons can be expressed as a linear equation where the parameter $D_{\rm ion}$ is the calibration relating the UV $1600{\rm \AA}$ luminosity to the ionising photon escape rate,
\begin{equation}
\dot{N}_{\rm ion\, pro}=D_{\rm ion}\times 10^{13}\,L_{1600}.
\end{equation}
Much like the UV luminosity to SFR calibration ($B_{\rm UV}$) $D_{\rm ion}$ is senstive to a number of factors including the previous star formation duration, the metallicity and the initial mass function. Assuming the same scenario as used in deriving the star formation rate density (i.e. $100\,{\rm Myr}$ previous star formation, $Z=0.02$, and a Salpeter IMF) the {\sc Starburst99} population synthesis model implies a value of $D_{\rm ion}=1.2$. Decreasing the previous star formation duration to $10\,{\rm Myr}$ and reducing the metallicity by a factor of $50$ yields results an increase in the calibration to $D_{\rm ion}=3.1$. Further, adopting a radically different IMF, such as truncating the formation of stars below $50\,M_{\odot}$ can produce a further increase by a factor of $\times 2$.

However, not all of the ionising photons are able to escape the immediate confines of the star forming region. Some fraction ($1-f_{\rm esc}$) are absorbed by the local ISM, leaving a reduced number that escape and are capable of ionising the surround neutral hydrogen,
\begin{equation}
\dot{N}_{\rm ion\, esc}=f_{\rm esc} \times \dot{N}_{\rm ion\, pro}
\end{equation}

To consider whether the Universe is ionisised we employ a modified version (to compensate for our choice of cosmology) of equation 26 of Madau, Hardt \& Rees (1999). This describes the number of ionising photons required to maintain the ionisation of the Universe as a function of redshift and the hydrogen clumping factor $C$,
\begin{equation}
\dot{N}_{\rm ion\, req}=2.51\times 10^{47}\,C\,(1+z)^{3}.
\end{equation}

\begin{figure}
\centering
\includegraphics[width=20pc]{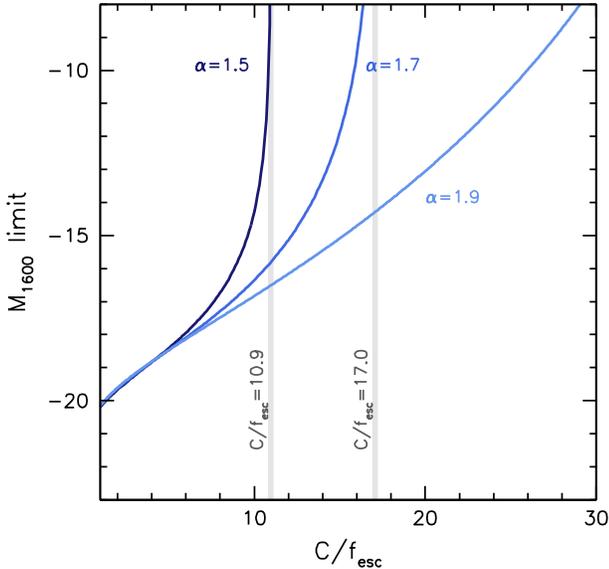}
\caption{The value of $M_{1600}$ it is necessary to integrate the best fit UV luminosity functions assuming fixed $\alpha\in\{1.5,1.7,1.9\}$ to maintain the ionisation of the Universe as a function of $C/f_{esc}$, i.e. the ratio of the clumping factor to the escape fraction of ionising photons.}
\label{fig:reionise_z}
\end{figure}

Given that we infer $\dot{N}_{\rm ion\, pro}$ from the UV luminosity density around $1600{\rm \AA}$ and have limited constraints on both $C$ and $f_{\rm esc}$ it is convenient to ascertain the range of values of $C/f_{\rm esc}$ for which the Universe can remain ionised assuming the luminosity functions for 3 different faint end slopes obtained in Section \ref{sec:lf}. Figure \ref{fig:reionise_z} shows the limit down to which the various luminosity functions need to be integrated to sustain the ionisation as a function of the parameter $C/f_{\rm esc}$. Assuming a lower integration limit of $M_{1600}=-18.5$, i.e. including only the contribution from galaxies brighter than our observed limit, $C/f_{\rm esc}$ is found be $<5.0$. Assuming a faint-end slope of $\alpha=-1.7$ and extrapolating to $M_{1600}<-8.0$ limits the value of $C/f_{\rm esc}$ to $<17.0$.

While the values of $C$ and $f_{\rm esc}$ remain uncertain a number of studies have attempted to constrain them through both observations and simulations. Early simulations suggested a value of $C\approx 30$ (Gnedin \& Ostriker 1997), while more recent attempts have yielded considerably smaller values (e.g. Pawlik et al. 2009: $C\approx 5$). The value of $f_{\rm esc}$ is similarly poorly constrained; recent observations (e.g. Siana et al. 2010) at $z\approx 1-2$ suggest a values of $\sim 0.05 - 0.20$, while simulations (at higher-redshift) suggest potentially larger escape fractions. Assuming $f_{\rm esc}=0.2$, and $C=5$ suggests $C/f_{\rm esc}=25$. This value is in excess of that capable for the directly observed population of galaxies (i.e. those with $M_{1600}<-18.5$) suggesting that we have not observed the full population responsible for reionisation. However, it is only mildly in excess of that possible if the luminosity function is extrapolated to fainter magnitudes assuming a faint-end slope of $\alpha=-1.7$. Slopes steeper than $\alpha=-1.86$ would, assuming there was a contribution from galaxies down to $M_{1600}=-8.0$, provide sufficent ionising photons to maintain the ionisation of the Universe at $z\approx 7$. Of course, reducing the metallicity or changing the IMF, would, for a given UV luminosity density increase the number of ionising photons. Thus, at $z\approx 7$, assuming $f_{\rm esc}>0.2$, and $C=5$ it appears that star forming galaxies could produce a sufficient number of ionising photons assuming a fairly steep faint-end slope and/or lower metallicity or different IMF.

\section{Conclusions}

In this work we have presented a catalogue of candidate $z$-drop $z\sim 7$ galaxies from 4 separate fields (HUDF, P34, P12, ERS) covering a total area of $\sim 50.0$ arcmin$^{2}$. The deepest field, the HUDF, extends to $J_{125w}=28.36$ ($7\sigma$) and shallowest, the ERS, $J_{125w}=27.03$ ($7\sigma$). In total we present 44 candidates (HUDF:11, P34:15, P12:7, ERS:11) covering an apparent magnitude range of $J_{125w}=25.9-28.36$. Simulations reveal these galaxies likely cover a redshift distribution $z=6.4-7.7$. Of these 44 candidates, 15 are recorded in a range of previous studies.

While our selection criteria guards against any intrinsic interloper population it is still susceptible to contamination through photometric scatter and transient phenomena (supernovae and significant apparent motion objects). Simulations testing the effect of photometric scatter suggest a contamination rate of $\sim 5\%$ ($\sim 2.5$ objects). Further, we note the presence of a possible supernovae in the HUDF field and record the presence of an object, which due to significant proper motion, mimics the characteristics of a high-redshift galaxy.

Using this new data set, we determine the stepwise rest-frame ultraviolet luminosity function for each field ($-21.0<M_{1600}<-18.5$ over the 4 fields). These estimations of the luminosity function are consistent with previous estimates at $z\approx 7$. Further, these appear to confirm that the observed decline in the bright-end of the UV luminosity function (and star formation rate density) with redshift seen between $z=3$ and $z=6$ continues to $z\approx 7$. 

The current data does not extend to sufficiently low luminosities to probe the faint-end slope of the luminosity function (assuming the Schechter parameterisation). However, by assuming various faint-end slopes $-1.7>\alpha>-1.9$, motivated by observations at lower redshift, the remaining luminosity function parameters can be determined. From these the ultraviolet luminosity density, and thus both the star formation rate and ionising photon luminosity densities can be estimated.

The ionising photon luminosity density inferred from the observed luminosity function extrapolated to $M_{1600}=-8.0$ assuming either a steep steep faint-end $\alpha<-1.86$ or a shallower slope with a lower metallicity and/or different IMF is sufficient to maintain the ionisation of the Universe for recent estimates of the hydrogen clumping factor and ionising photon escape fraction.

\subsection*{Acknowledgements}
We would like to thank the referee for their detailed and helpful suggestions. SMW acknowledges the support of STFC and SL and JC acknowledge the support of ELIXIR. Based on observations made with the NASA/ESA Hubble Space Telescope,
obtained from the Data Archive at the Space Telescope Science Institute, which is operated by the Association
of Universities for Research in Astronomy, Inc., under NASA contract
NAS 5-26555. These observations are associated with programme \#GO/DD-11359 and \#GO-11563. We are grateful to
the  WFC\,3 Science Oversight Committee for making their Early Release Science observations public.

\bsp


\begin{thebibliography}{}
\bibitem[Beckwith et al.(2006)]{2006AJ....132.1729B} Beckwith, S.~V.~W., et al.\ 2006, AJ, 132, 1729

\bibitem[\protect\citename{Bertin \& Arnouts} 1996]{be96}
Bertin E., Arnouts S., 1996, A\& AS, 117, 393

\bibitem[Bouwens et al.(2004)]{2004ApJ...616L..79B} 
Bouwens, R.~J., et al.\ 2004, \apjl, 616, L79 

\bibitem[\protect\citename{Bouwens et al.\ } 2006]{bo06}
Bouwens R.~J. Illingworth G.~D. Blakeslee J.~P.; Franx M., 2006, ApJ, 653, 53

\bibitem[\protect\citename{Bouwens et al.\ } 2008]{bo08}
Bouwens R.~J., Illingworth G.~D., Franx M., Ford H., 2008, ApJ, 686, 230

\bibitem[\protect\citename{Bouwens et al.\ } 2009]{bo09}
Bouwens R.~J. et al., 2009, submitted to ApJ letters, arXiv:0909.1803

\bibitem[\protect\citename{Bremer et al.} 2004]{br04}
Bremer M.~N., Lehnert M.~D., Waddington I., Hardcastle M.~J., Boyce P.~J., Phillipps S., 2004, MNRAS, 347, L7

\bibitem[Bunker et al.(2003)]{2003MNRAS.342L..47B} Bunker, A.~J., Stanway, 
E.~R., Ellis, R.~S., McMahon, R.~G., 
\& McCarthy, P.~J.\ 2003, \mnras, 342, L47 

\bibitem[\protect\citename{Bunker et al.\ } 2004]{bu04}
Bunker A.~J., Stanway E.~R., Ellis R.~S., McMahon R.~G., 2004, MNRAS,
355, 374

\bibitem[Bunker et al.(2009)]{2009arXiv0909.2255B} Bunker, A., et al.\ 
2009, arXiv:0909.2255 





\bibitem[Fan et al.(2001)]{2001AJ....122.2833F} Fan, X., et al.\ 2001, \aj, 
122, 2833 

\bibitem[\protect\citename{Ferguson et al.\ } 2004]{fe04}
Ferguson H,~C. et al., 2004, ApJ, 600, 107

\bibitem[Finkelstein et al.(2009)]{2009arXiv0912.1338F} Finkelstein, S.~L., 
Papovich, C., Giavalisco, M., Reddy, N.~A., Ferguson, H.~C., Koekemoer, 
A.~M., \& Dickinson, M.\ 2009, arXiv:0912.1338 

\bibitem[Giavalisco et al.(2004)]{2004ApJ...600L.103G} Giavalisco, M., et 
al.\ 2004, ApJ, 600, L103 


\bibitem[Gunn 
\& Peterson(1965)]{1965ApJ...142.1633G} Gunn, J.~E., \& Peterson, B.~A.\ 1965, \apj, 142, 1633 



\bibitem[Koekemoer et al.(2002)]{2002hstc.conf..337K} Koekemoer, A.~M., Fruchter, A.~S., Hook, R.~N., \& Hack, W.\ 2002, The 2002 HST Calibration Workshop : Hubble after the Installation of the ACS and the NICMOS Cooling System, 337

\bibitem[\protect\citename{Knapp et al.\ } 2004]{kn04}
Knapp G.~R. et al., 2004, AJ, 127, 3553


\bibitem[Leitherer et al.(1999)]{1999ApJS..123....3L} Leitherer, C., et 
al.\ 1999, \apjs, 123, 3 


\bibitem[McLure et al.(2009)]{2009arXiv0909.2437M} McLure, R.~J., Dunlop, 
J.~S., Cirasuolo, M., Koekemoer, A.~M., Sabbi, E., Stark, D.~P., Targett, 
T.~A., \& Ellis, R.~S.\ 2009, arXiv:0909.2437 


\bibitem[\protect\citename{Oke \& Gunn} 1983]{ok83}
Oke J.~B., Gunn J.~E., 1983, ApJ, 266, 713

\bibitem[Oesch et al.(2009)]{2009arXiv0909.1806O} Oesch, P.~A., et al.\ 
2009, arXiv:0909.1806 

\bibitem[Oesch et al.(2007)]{2007ApJ...671.1212O} Oesch, P.~A., et al.\ 2007, ApJ, 671, 1212


\bibitem[\protect\citename{Ouchi et al.\ } 2009]{ou09}
Ouchi M. et al., 2009, arXiv, 0908, 3191

\bibitem[Ota et al.(2008)]{2008ApJ...677...12O} Ota, K., et al.\ 2008, 
ApJ, 677, 12 


\bibitem[Reddy 
\& Steidel(2009)]{2009ApJ...692..778R} Reddy, N.~A., \& Steidel, C.~C.\ 2009, ApJ, 692, 778 




\bibitem[Schlegel et al.(1998)]{1998ApJ...500..525S} Schlegel, D.~J., 
Finkbeiner, D.~P., \& Davis, M.\ 1998, \apj, 500, 525 

\bibitem[\protect\citename{Stanway, Bunker \& McMahon} 2003]{st03}
Stanway E.~R., Bunker A.~J., McMahon R.~G., 2003, MNRAS, 342, 439

\bibitem[Stanway et al.(2004)]{2004ApJ...604L..13S} Stanway, E.~R., et al.\ 
2004, \apjl, 604, L13 

\bibitem[Stanway et al.(2004)]{2004ApJ...607..704S} Stanway, E.~R., Bunker, 
A.~J., McMahon, R.~G., Ellis, R.~S., Treu, T., 
\& McCarthy, P.~J.\ 2004, \apj, 607, 704 

\bibitem[\protect\citename{Stanway, McMahon \& Bunker} 2005]{st05}
Stanway E.~R., McMahon R.~G., Bunker A.~J., 2005, MNRAS, 359, 1184


\bibitem[\protect\citename{Steidel et al.\ } 1996]{st96}
Steidel C.~C., Giavalisco M., Pettini M., Dickinson M., Adelberger K.~L., 1996, ApJ, 462, 17

\bibitem[Steidel et al.(1999)]{1999ApJ...519....1S} Steidel, C.~C., 
Adelberger, K.~L., Giavalisco, M., Dickinson, M., 
\& Pettini, M.\ 1999, \apj, 519, 1 




\bibitem[Wilkins et al.(2010)]{2010MNRAS.tmp..176W} Wilkins, S.~M., Bunker, 
A.~J., Ellis, R.~S., Stark, D., Stanway, E.~R., Chiu, K., Lorenzoni, S., 
\& Jarvis, M.~J.\ 2010, \mnras, 176 

\bibitem[Yan et al.(2009)]{2009arXiv0910.0077Y} Yan, H., Windhorst, R., 
Hathi, N., Cohen, S., Ryan, R., O'Connell, R., 
\& McCarthy, P.\ 2009, arXiv:0910.0077 








\end{thebibliography}
\end{document}